\documentclass[useAMS,usenatbib]{mnras}

\usepackage{graphicx}
\usepackage{rotating}
\usepackage{caption}
\usepackage{subcaption}

\usepackage{romannum}
\usepackage{color}
\usepackage{amsmath}
\usepackage{multirow}

\newlength{\colwidth}
\newcommand{\red}{0.495}
\newcommand{\sred}{0.46}

\newcommand{\Msol}{\, \rm M_{\odot}}
\newcommand{\hMsol}{\, h^{-1}{\rm M_{\odot}}}
\newcommand{\hMpc}{\, h^{-1}{\rm Mpc}}

\newcommand{\hpc}{\, h^{-1}{\rm pc}}
\newcommand{\gyr}{\, {\rm Gyr}}
\newcommand{\myr}{\, {\rm Myr}}

\newcommand{\hnMsol}{\, h_{70}^{-1}{\rm M_{\odot}}}

\newcommand{\Mpc}{\, {\rm Mpc}}

\newcommand{\SFR}{{\rm SFR}}
\newcommand{\sSFR}{{\rm sSFR}}

\newcommand{\kms}{{\rm km}\, {\rm s}^{-1}}

\newcommand{\HI}{\textrm{H}\,\textsc{i}}

\newcommand{\mstarpivot}{{$10^{9} \Msol \,$}}
\newcommand{\mstarmin}{{$10^{8} \Msol \,$}}

\newcommand{\res}{7.964357}

\renewcommand{\tabcolsep}{1pt}

\begin{document}

\title[DRAGONS IX: Economics of reionizing galaxies]{Dark-ages reionization and galaxy formation simulation - \Romannum{9}. Economics of reionizing galaxies}

\author[A. R. Duffy et al.]
{Alan R. Duffy$^{1}$, Simon J. Mutch$^{2}$, Gregory B. Poole$^{2}$, Paul M. Geil$^{2}$,  \newauthor Han-Seek Kim$^{2}$, Andrei Mesinger$^{3}$ and J. Stuart B. Wyithe$^{2}$ \\
$^1$Centre for Astrophysics and Supercomputing, Swinburne University of Technology, Hawthorn, VIC 3122, Australia, \\
$^2$School of Physics, University of Melbourne, Parkville, VIC 3010, Australia, \\
$^3$Scuola Normale Superiore, Piazza dei Cavalieri 7, I-56126 Pisa, Italy}

\date{}

\maketitle

\label{firstpage}

\begin{abstract}
Using a series of high-resolution hydrodynamical simulations we show that during the rapid growth of high-redshift ($z > 5$) galaxies,
reserves of molecular gas are consumed over a time-scale of $300 \myr$ almost independent of feedback scheme. 
We find that there exists no such simple relation for the {\it total} gas fractions of these galaxies, with little correlation between gas fractions 
and specific star formation rates.
The bottleneck or limiting factor in the growth of early galaxies is in converting infalling gas to cold star-forming gas.
Thus, we find that the majority of high-redshift dwarf galaxies are effectively in recession, with demand (of star formation) never rising to 
meet supply (of gas), irrespective of the baryonic feedback physics modelled. 
We conclude that the basic assumption of self-regulation in galaxies -- that they can adjust total gas consumption within a Hubble time -- 
does not apply for the dwarf galaxies thought to be responsible for providing most UV photons to reionize the high-redshift Universe.
We demonstrate how this rapid molecular time-scale improves agreement between semi-analytic model predictions of the early Universe
and observed stellar mass functions. 
\end{abstract}


\section{Introduction}
\label{Introduction}
The study of galaxy formation at high-redshifts, $z>5$, has undergone a revolution in the last several years with deep {\it Hubble Space Telescope}
images, primarily the {\it Hubble Ultra and eXtreme Deep Fields}~\citep{Beckwith:06,Illingworth:13}, enabling the study of 
galaxy populations in the first $\gyr$ after the big bang. Additionally, new radio/submillimetre facilities such as the Karl G. Jansky Very Large Array, 
Atacama Large Millimeter/submillimeter Array and the 
Institut de Radio Astronomie Millimetrique Plateau de Bure Interferometer have provided resolved gas phase information at high-redshifts. Alongside the increase 
in observational data sets is a commensurate increase in theoretical efforts to explain these early galaxies' high star formation
rates~\citep[SFRs; e.g.][]{Daddi:07,Noeske:07,Franx:08} and large gas fractions~\citep[e.g.][]{Daddi:10,Genzel:10,Tacconi:10,Casey:11}.

A general framework to explore galaxy growth is termed `baryon cycling', whereby matter and energy flow into and out of a galaxy
and the surrounding intergalactic medium. It has been argued through hydrodynamic simulations~\citep[e.g.][]{Finlator:08,Dave:12} 
as well as scaling relations~\citep[e.g.][]{Bouche:10,Dutton:10} that inflows and outflows will tend to balance, provided the 
SFR of a galaxy can adjust sufficiently quickly. 

As discussed in the toy model of~\citet{Schaye:15}, an increase in the galaxy inflow rate, $\dot{M}_{\rm inflow}$ leads to a growth in the 
gas reservoir, $\dot{M}_{\rm gas}$, which can then form more stars $\dot{M}_{\rm \star}$. These stars then explode
increasing the outflow of material $\dot{M}_{\rm outflow}$ as gas in the supernovae-driven winds. This interplay is described in
\begin{equation}\label{eqn:equilibrium}
\dot{M}_{\rm inflow} = \dot{M}_{\rm outflow} + \dot{M}_{\rm \star} + \dot{M}_{\rm gas} \,,
\end{equation}
The reverse is then also true. If inflow rates are less than the outflow rates, the gas reservoir decreases and star formation (SF) will begin 
to slow, thereby decreasing the resultant outflow rate. Ultimately, averaged over appropriate time-scales, outflows will balance inflows.
In this regime galaxies are described by an \emph{equilibrium model}~\citep[e.g.][]{WhiteFrenk:91,Schaye:10,Lilly:13}, 
also called the `bathtub model'~\citep{Bouche:10,Dave:12}. Another way to phrase this bathtub model is that galaxies are `supply-side'
limited and grow according to input gas flows.

If the system is in equilibrium then the gas reservoir changes ($\dot{M}_{\rm gas}$) are negligible~\citep{Schaye:15} and thus 
$\dot{M}_{\rm \star}  = \dot{M}_{\rm inflow} - \dot{M}_{\rm outflow}$. 
The resultant gas reservoir level in this picture does not directly depend on the feedback strength but rather is a consequence of the SFR
that is needed to balance the inflows with outflows, provided the gas consumption time-scale is less than a Hubble time~\citep{Schaye:15}.
As shown by~\citet{Krumholz:12b}, depending on the scaling relation used, star-forming galaxies at even  
modest redshifts ($z>2$) may not achieve this equilibrium state.

In this work, we will explore the ideas of gas consumption and replenishment at high-redshift during the Epoch of Reionization. 
The shorter Hubble time in this epoch may result in most galaxies being out of equilibrium. In particular, the gas reservoir term, $\dot{M}_{\rm gas}$, may
not be negligible. Simple scaling arguments suggest that this may in fact be unavoidable at sufficiently high-redshift.
In the (dense) early Universe galaxies can achieve high SFRs due to short dynamical times ($t_{\rm dyn} \propto t_{H} \propto (1+z)^{-1.5}$).
However, the gas may be even more rapidly replenished thanks to high infall rates, $\dot{M}_{\rm inflow}$, which scale
with cosmic infall as $(1+z)^{2.25}$~\citep[e.g.][]{Dekel:09,Correa:15a,Correa:15b,Correa:15c}.
The situation where local demand (SFR) cannot respond to increasing supply (gas) is termed a recession in economics. 

We aim to explore early galaxy growth using high-resolution hydrodynamical simulations with various
feedback schemes, determining the conditions under which early galaxies are supply-side limited. 
Studies of reionization require the physics of low-mass galaxy formation to be extended 
over volumes of $(100s \Mpc)^3$. While these large volume hydrodynamical simulations are becoming increasingly
available~\citep[e.g.][]{Vogelsberger:14a,Vogelsberger:14b,Schaye:15}, they remain extremely computationally expensive.
Traditionally, semi-analytic models (SAMs) have been used to explore the space of the various physical parameters in high-redshift galaxy 
formation~\citep[e.g.][]{Benson:06,Lacey:11,Raicevic:11,Zhou:13}.
These SAMs are computationally much cheaper and run as a post-processing step on an existing $N$-body simulation,
which can greatly exceed the hydrodynamical simulation in dynamic mass range probed for a comparable computational costs.
However, in this work we hope to offer insight into emergent behaviour from the hydrodynamic simulations that can be 
incorporated into SAMs that study the Epoch of Reionization and thereby reduce their parameter space.

In Section~\ref{sec:simulations} we describe the \textsc{Smaug} hydrodynamical simulations used in this work, and then explore the 
rate at which gas is consumed in the early galaxies in Section~\ref{sec:gas_frac}. We explore the dependence
of the consumption time-scales on stellar mass and redshift in Section~\ref{sec:f_smass}.  
The gas distribution as a function of mass is shown in Section~\ref{sec:gas_dist}, demonstrating that the vast majority of 
systems are indeed gas rich at high-redshifts. We then summarize our findings in Section~\ref{sec:results}, and discuss the implications
of these findings for SAMs of early galaxy formation. 

\section{Simulation details}
\label{sec:simulations}
The \textsc{Smaug} simulation is a series of high-resolution hydrodynamical runs of a cosmological volume created as part of the 
Dark-ages Reionization And Galaxy Observables from Numerical Simulations (DRAGONS) project.\footnote{http://dragons.ph.unimelb.edu.au/}
DRAGONS further includes the large $N$-body \textsc{Tiamat} simulation~\citep{Poole:16} used to analyse large-scale dark matter (DM) structures 
during the era of first galaxy formation~\citep{Angel:16}. The DM haloes then have galaxy properties applied with a bespoke SAM, \textsc{meraxes}~\citep{Mutch:16a},
which includes coupled photoionization feedback through an excursion set scheme \textsc{21cmfast}~\citep{Mesinger:11} to track reionization. 
\textsc{Meraxes} has successfully reproduced existing observations at high-redshifts ($5<z<10$) of the luminosity and stellar mass functions~\citep{Liu:16}, sizes of 
galaxies~\citep{Liu:17} as well as predicted the reionization structure of the intergalactic medium around the growing galaxies~\citep{Geil:16}.

There are a number of free parameters in the \textsc{meraxes} SAM. In this work we have investigated whether
it is possible to use the emergent properties and scaling relationships from the explicitly modelled infall and outflow of the gas 
hydrodynamics in \textsc{Smaug} to inform the semi-analytics. In particular, we investigate
the overall halo gas consumption times in early galaxies and demonstrate that gas consumption rates can be fixed independently of supernovae (SNe) feedback.
This is potentially a substantial reduction in the freedom for early Universe models and ensures that maximum physical insight can be 
gained by comparing SAMs with the limited observations currently available.

\begin{figure}
\includegraphics[width=\colwidth,keepaspectratio=true]{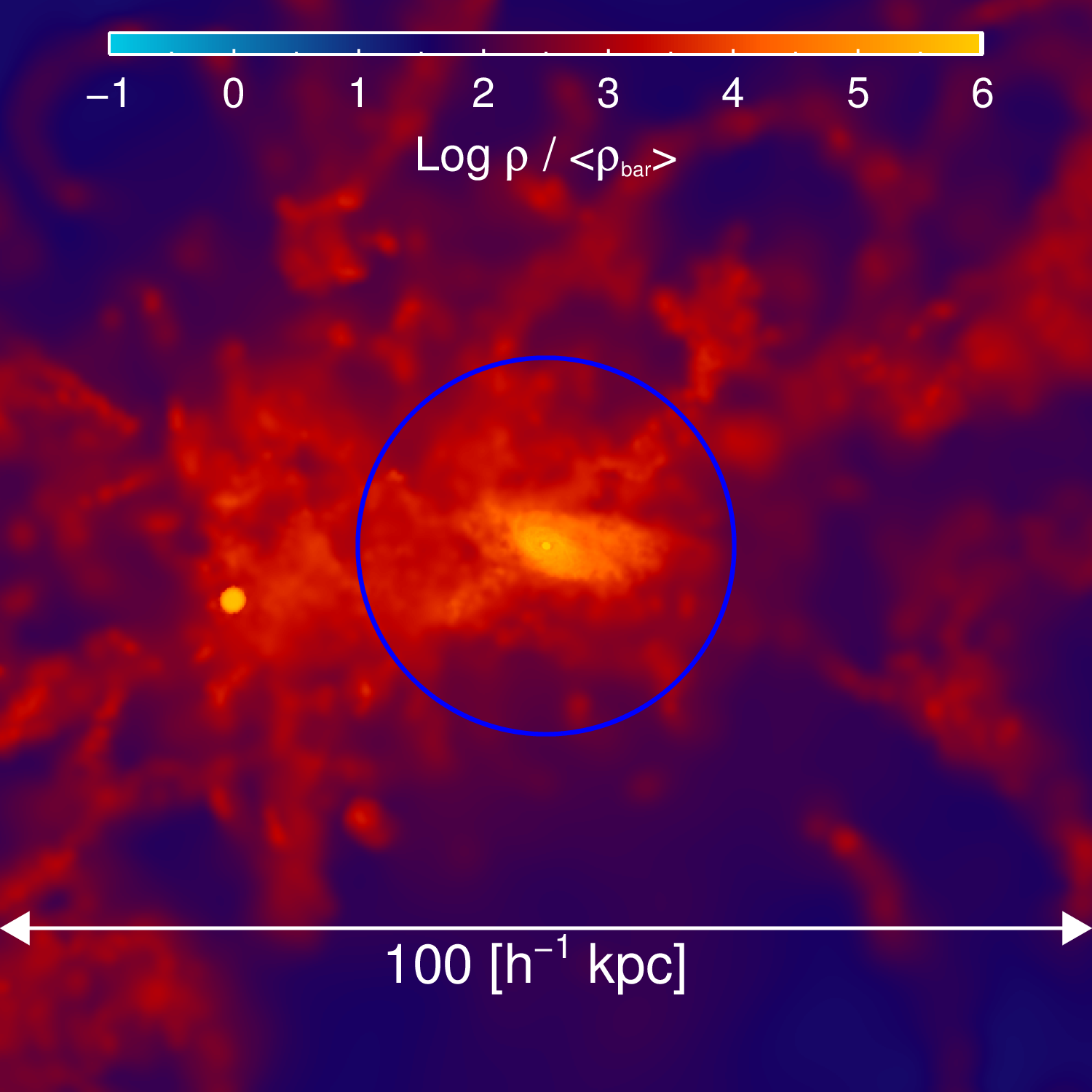} 
\caption{The gas density in a forming galaxy at $z=5$ simulated with the maximal SNe feedback physics (\emph{ZC\_SSNe\_Thermal}) in the \textsc{Smaug} series. 
A distinct disc has formed and a smaller satellite is inspiralling, likely to merge within $100 \myr$. 
It is important to note the significant amount of structure around the disc, showing a large amount of material in/out-flowing that complicates the modelling of a galaxies at high-redshifts.}\label{fig:galpic}
\end{figure}

\begin{table*}
\begin{center}
\caption{A list of the simulated galaxy formation physics schemes utilized in this study. From left to right, the 
columns list the simulation name used in this study; the name as defined by~\citet{Schaye:10} (where appropriate, else `--' used to indicate new);
and a brief description of the physics modelled. Simulations with \emph{LateRe} (\emph{EarlyRe}) have instantaneous reionization
occur at $z=6.5$ $(9)$ given by a~\citet{HaardtMadau:01} UV/X-ray background thereafter.}
\label{tab:physics}
\begin{tabular}{lll}
\hline
Simulation & OWLS name & Brief description\\
\hline
\emph{PrimC\_NoSNe} & \emph{NOSN\_NOZCOOL} & No energy feedback from SNe and  cooling assumes primordial abundances\\
\emph{PrimC\_WSNe\_Kinetic} & \emph{NOZCOOL}  & Cooling as \emph{PrimC\_NoSNe} but with fixed mass loading of `kinetic' winds from SNe \\
\emph{ZC\_WSNe\_Kinetic} & \emph{REF} & SNe feedback as \emph{PrimC\_WSNe} but now cooling rates include metal-line emission \\
\emph{ZC\_SSNe\_Kinetic} & \emph{--} & Strong SNe feedback as \emph{ZC\_WSNe\_Kinetic} but using 100 per cent of available SNe energy \\
\emph{ZC\_SSNe\_Thermal} & \emph{WTHERMAL} & Strong SNe feedback as \emph{ZC\_SSNe\_Kinetic} but feedback modelled thermally \\
\hline
\end{tabular}
\end{center}
\end{table*}

\subsection{Galaxy physics}
The \textsc{Smaug} simulations~\citep{Duffy:14b} were run with a version of \textsc{gadget3}, the $N$-body/hydrodynamic
code described in~\citet{Springel2005b}, with new modules for radiative cooling~\citep{Wiersma:09a}, SF~\citep{Schaye:08},
stellar evolution and mass-loss~\citep{Wiersma:09b}, and galactic winds driven kinetically by SNe as given in~\citet{DallaVecchia:08}, or thermally SNe driven winds as implemented
in~\citet{DallaVecchia:12}. 
The \textsc{Smaug} series of simulations all share the same initial conditions but are resimulated using numerous different physics models. These models are 
a subset of those tested in the OverWhelmingly Large Simulations series (OWLS;~\citealt{Schaye:10}) 
which allows for robust investigation of the relative importance of each scheme, without exploring irrelevant physics to the high-redshift Universe (e.g. differing Type Ia SN models).
These physics models have had success in reproducing observed galaxy properties at low, post-reionization, redshifts.
The strong feedback schemes tested here have had success in reproducing the cosmic SF history~\citep{Schaye:10}, 
as well as the high-mass end of the \HI\ mass function~\citep{Duffy:12a} at $z<2$. Also, certain schemes reproduced the stellar mass function and specific star formation 
rates (sSFRs) at $z=0-2$~\citep{Haas:13a,Haas:13b}.
We briefly cover the simulation details here and summarize them in Table~\ref{tab:physics}, but refer the reader to~\citet{Duffy:10} and~\citet{Schaye:10} for more explicit explanations
of the physics, and~\citet{Duffy:14b} for greater detail of the \textsc{Smaug} simulations.

\subsection{Initial conditions}
Each simulation followed $N^3$ DM particles, and $N^3$ gas particles, where $N = 512$, within periodic cubic volumes of comoving length 
$L = 10\hMpc$. The Plummer-equivalent comoving softening length is $200 \hpc$ and the DM (gas/star) particle mass 
is $4.7\, (0.9) \times 10^{5} \hMsol$.
Lower resolution test cases with $N=128$ and $256$ were also run (see the Appendix).
For all cases, we used grid-based cosmological initial conditions  generated with \textsc{grafic}~\citep{Bertschinger:01} at 
$z=199$ using the Zeldovich approximation and a transfer function from \textsc{cosmics}~\citep{Bertschinger:95}.

We use the {\it Wilkinson Microwave Anisotropy Probe} 7 year results~\citep{Komatsu:11}, henceforth known as \textit{WMAP}7, to set the cosmological parameters 
$[\Omega_{\rm m},$ $\Omega_{\rm b},$ $\Omega_{\Lambda},$ $h,$ $\sigma_{8},$ $n_{\rm s}]$ to [0.275, 0.0458, 0.725, 0.702, 0.816, 0.968] and $f^{\rm univ}_{\rm b} = 0.167$. 

After resolution testing (see the Appendix), we adopt a conservative cut on halo mass of $M_{\rm halo, \rm lim}=10^{8.5} \hMsol$, corresponding to $>650$ DM particles, or on stellar mass of $M_{\rm \star, \rm lim}=10^{6.7} \hMsol$ ($>50$ star particles) depending on the quantity explored. Note that all resolution limits are indicated on the plots used.

\subsubsection{Radiative cooling}
We follow the radiative cooling (and heating) prescription of~\citet{Wiersma:09a} in which cooling (and heating) rates are assigned to each gas particle based on interpolated density 
and temperature tables from \textsc{cloudy}~\citep{Cloudy}. We track net radiative cooling rate for elements in the presence of the cosmic microwave background 
and (after reionization) a~\citet{HaardtMadau:01} UV / X-ray background from quasars and galaxies. 
For our models named \emph{PrimC} we only track this emission-line cooling from hydrogen and helium, while those named \emph{ZC} have additional contributions
from carbon, nitrogen, oxygen, neon, magnesium, silicon, sulphur, calcium and iron. The metallicity of each particle is calculated by weighting the nearest 
 $N_{\rm ngb}=48$ neighbours according to the smoothing kernel~\citep{Wiersma:09b}. 
At high temperatures $\gg10^{5}K$ (seldom reached in \textsc{Smaug} for $5<z<20$) emission of bremsstrahlung radiation or inverse Compton scattering off cosmic microwave 
background photons dominates ionized gas cooling rates. Note that we do not model the expected molecular cooling from our simulations and instead calculate
the molecular fractions in a post-processing scheme discussed in Section~\ref{sec:molgas}.

Note that we have modelled reionization as a background that is `switched on' at $z=9$ ($6.5$) for simulations with \emph{EarlyRe-} (\emph{LateRe-}) ionization epochs.
Such an instantaneous reionization is a relatively good approximation as reionized bubbles are comparable in scale~\citep{Wyithe:04} to the simulation box sizes in \textsc{Smaug}. 
However, we use an optically thin approximation in which all gas particles are exposed to the background field. This will increase the fraction of gas that is impacted by photoionization 
feedback (ignoring self-shielding of dense gas) but underestimate the temperature immediately after reionization~\citep[e.g][]{Wiersma:09a,Schaye:10}. We note that local stellar 
sources of ionizing photons form in those regions that would be identified as self-shielded, thereby adding extra photons locally and in part mitigating this shielding. This mitigation 
means that the final neutral gas density distribution lies between the optically thin case and a cruder self-shielding approximation that lacks coupled radiative transfer, as shown in~\citet{Rahmati:13b}.

\subsubsection{Star formation}
As gas cools, its density rises until reaching a critical point (with number density $n_{H} = 10^{-2}-10^{-1} {\rm cm^{-3}}$) where it experiences 
instabilities leading to a multiphase medium and ultimately the formation of stars~\citep{Schaye:04}. Our cosmological simulation lacks
the resolution (as well as physics) to model molecular hydrogen formation and SF directly in multiphase regions. 
Therefore, as in~\citet{Schaye:10} we have implemented\footnote{By setting $\gamma_{\rm eff} = 4/3$ we ensure that the Jeans mass
and ratio of the Jeans length to the SPH smoothing kernel are independent of density (as demonstrated in~\citealt{Schaye:08}).} an effective equation of state $\gamma_{\rm eff}$
with pressure $P \propto \rho^{\gamma_{\rm eff}}$ for gas above a critical hydrogen number density, $n^{*}_{H} = 0.1 {\rm cm^{-3}}$.
These star-forming gas particles are converted to stars according to a pressure-based star formation SF law, which was shown by~\citet{Schaye:08} to 
reproduce the observed Kennicutt--Schmidt law~\citep{Kennicutt:98b}. During SF, we convert one gas particle to one stellar particle
as `spawning' multiple generations of stars from a given gas particle leads to numerical artefacts and reduced effectiveness of SF feedback~\citep{DallaVecchia:08}. 

We will explore the consequences of modifying the SF law (in particular with a metallicity-dependent threshold suggested in~\citealt{Schaye:04}) 
in a future publication, and emphasize that our conclusions in this work are restricted to an SF law with a fixed density threshold and parameter choice for the power law that reproduces the locally observed Kennicutt--Schmidt law.

\subsubsection{Molecular hydrogen estimation}\label{sec:molgas}
As previously noted, we do not have the required resolution to track the multiphase interstellar medium (ISM) within the cosmological volume of \textsc{Smaug}. Instead we use 
the empirical formalism described in~\citet{Duffy:12a} to estimate the molecular hydrogen fractions based on the pressure of the dense star-forming 
equation-of-state gas. We briefly describe this methodology here but refer the reader to~\citet{Duffy:12a} for more information. 

We adopt the empirical scaling of the THINGS survey~\citep{Leroy:08} of a power-law relation between the surface density ratio of molecular-to-atomic hydrogen gas, $R_{\rm surf}$, and the ISM pressure $P$:
\begin{equation}\label{eqn:rsurf}
R_{\rm surf} =  \left( \frac{P/k}{10^{4.23} \,{\rm K cm^{-3}}} \right)^{0.8}\,.
\end{equation}
By assuming that hydrogen in star-forming gas is essentially atomic neutral and molecular phases (with a negligible mass fraction of ionized hydrogen), this
ratio then results in 
\begin{equation}\label{eqn:molmass}
M_{\rm H2} = M_{\rm gas}R_{\rm surf}/(1 + R_{\rm surf})\,.
\end{equation}
In practice, most of the star-forming gas at high-redshift is at such high pressures that the molecular component completely dominates the atomic neutral
hydrogen fraction.

The pressure used in equation (\ref{eqn:rsurf}) is averaged over lengthscales of 800 and 400 pc for local galaxies and dwarf galaxies in~\citet{Leroy:08}, respectively. This
is well above our resolution limit in \textsc{Smaug}, ensuring that these regions are well sampled. We note however that the empirical result is a {\it local} relation 
and therefore extending this to high-redshifts is potentially dangerous. More involved molecular hydrogen formation schemes, such as~\citet{Gnedin:11} 
and~\citet{Krumholz:13} have been tested in~\citet{Lagos:15} against the~\citet{Duffy:12a} prescription. Overall,~\citet{Lagos:15} found that 82 (99) per cent of 
the molecular hydrogen identified by the~\citet{Gnedin:11}; \citet{Krumholz:13} model was found in star-forming gas particles
that~\citet{Duffy:12a}, and hence this work, would by definition expect to contain molecular hydrogen. 

The similarity between these schemes gives us confidence to continue with our simpler, empirically motivated prescription. We will explore more 
physically motivated schemes in future work, as~\citet{Lagos:15} found that significant molecular hydrogen reserves can be found in non-star-forming gas 
as a result of more advanced molecular gas formation schemes (e.g.~\citealt{Gnedin:11} and \citealt{Krumholz:13}) that depend on metallicity. In particular, 
even though we track metals in \textsc{Smaug} do not model the formation of molecular hydrogen on the expected dust grains from these metals and defer
this for a subsequent simulation set.

\subsubsection{Stellar evolution}
We assume that each star formed is a single stellar population from a~\citet{Chabrier:03} initial mass function. Within the star particle we track the formation and timed
release of metals from massive stars, as well as feedback (and chemical enrichment) from SNe Types Ia and II as described in~\citet{Schaye:10}. For the no-feedback 
model \emph{NoSNe} we do not couple the SNe feedback to the gas as a test case. In the other simulations, after a delay
time of $30 \rm Myr$, we assume that all massive stars (initial mass range $6-100 \Msol$) have ended as core-collapse SNe 
releasing $10^{51} \rm erg$ of energy. In this work we couple that resultant energy to the surrounding gas particles using one of the two techniques described below.

The first is a kinetic feedback scheme in which the nearest $N_{\rm ngb}=48$ neighbouring gas particles $i$ of a newly formed star $j$ have a 
probability of receiving a `kick' of wind velocity $v_{w}$ given by $\eta m_{j} / \Sigma^{N_{\rm ngb}}_{i}m_{i}$, where $\eta$ is the so-called mass-loading variable. 
For our weak feedback kinetic model \emph{WSNe\_Kinetic}, we adopt $\eta = 2$ and $v_{w}=600 \kms$ that corresponds to $f_{\rm SNe} = 0.4$ 
of the available SNe energy being coupled into the wind. As was noted in~\citet{Schaye:10}, the value
of $40$ per cent is reasonable as we do not model radiative losses in the \emph{Kinetic} scheme. 
We also have tested the feedback scheme of~\citet{DallaVecchia:12}, which injects the energy stochastically into 
a neighbouring gas particle, and in principle accounts for all the energy of the SN explosion $f_{\rm SNe} = 1$, a model we call \emph{SSNe\_Thermal}. To disentangle the impact that
the additional energy of the explosion has in \emph{SSNe\_Thermal} with the different feedback implementation of \emph{Kinetic} we ran a `strong' kinetic model \emph{SSNe\_Kinetic}.
This strong kinetic scheme has mass-loading $\eta = 3$ and wind velocity $v_{w}=774  \kms$ that corresponds to all of the SNe energy, $f_{\rm SNe} = 1$, coupled into the wind.

It is worth noting that we inject energy locally and never hydrodynamically decouple gas particles driven outwards in the wind. As shown by~\citet{DallaVecchia:08}, this is of key importance
in creating realistic disc structures.
An example disc galaxy from this particular simulation scheme is given in Fig.~\ref{fig:galpic}, showing the complexity of the 
gas distribution around newly forming systems that typically are dynamically unrelaxed at high-redshifts~\citep{Poole:16}.

\subsection{Power law}
In this work, we fit an evolving power law to several quantities of interest (denoted here as $\rm Var$), with the following functional form
\begin{equation}\label{eqn:evopowerlaw}
{\rm Var} = 10^{A_{\rm Var}} \left(\frac{M_{X}}{ M_{\rm pivot}} \right)^{B_{\rm mass}}\left(\frac{1+z}{1+z_{\rm pivot}} \right)^{C_{\rm evo}} \,,
\end{equation}
where $M_{X}$ is either the stellar (or total halo) mass, with corresponding pivot mass $M_{\rm pivot} = 1\, (5) \times 10^{9}\hnMsol$, for a given redshift $z$, normalized
at $z_{\rm pivot}=7$. We explicitly state those situations when we restricted the stellar mass fitting range to be above \mstarmin as it is the high-mass end which is
often best described by this power law. In practice, we take the quantity of interest over the redshift range of $5-10$ (in redshift intervals of $0.1$) and bootstrap the sample at each redshift
creating $5000$ samples that are binned across redshift with a uniform weighting applied during the least-squares (Levenberg--Marquardt) fit. We then record the solution to the bootstrap
ensembles to estimate the confidence intervals around the best fit to the full sample.

\section{Gas fractions}\label{sec:gas_frac}
We now consider the predictions that equilibrium models, such as~\citet{Bouche:10},~\citet{Dutton:10} and~\citet{Dave:12}, suggest should exist between
basic galaxy properties, in particular that early galaxies are more gas rich than low-redshift counterparts.
The initial equilibrium relation, equation (\ref{eqn:equilibrium}), describes the link between SF rising to consume inflows and driving outflows
to achieve balance in the consumption of the primary raw material in a galactic economic system, that is, gas. We can separate the molecular gas fraction 
(tracing a denser star-forming phase of the gas) into two SF-related quantities 
\begin{equation}\label{eqn:fhtwo}
f_{\rm H2} \equiv M_{\rm H2} / (M_{\rm H2} + M_{\star}) = \frac{1}{1 + (\tau_{\rm H2} \, \sSFR)^{-1}} \,,
\end{equation}
with gas consumption time-scale $\tau_{\rm H2}  = M_{\rm H2} / \SFR$ and specific star formation rate $\sSFR=\SFR/M_{\star}$. The molecular gas
mass estimation is described in Section~\ref{sec:molgas}.

\begin{figure*}
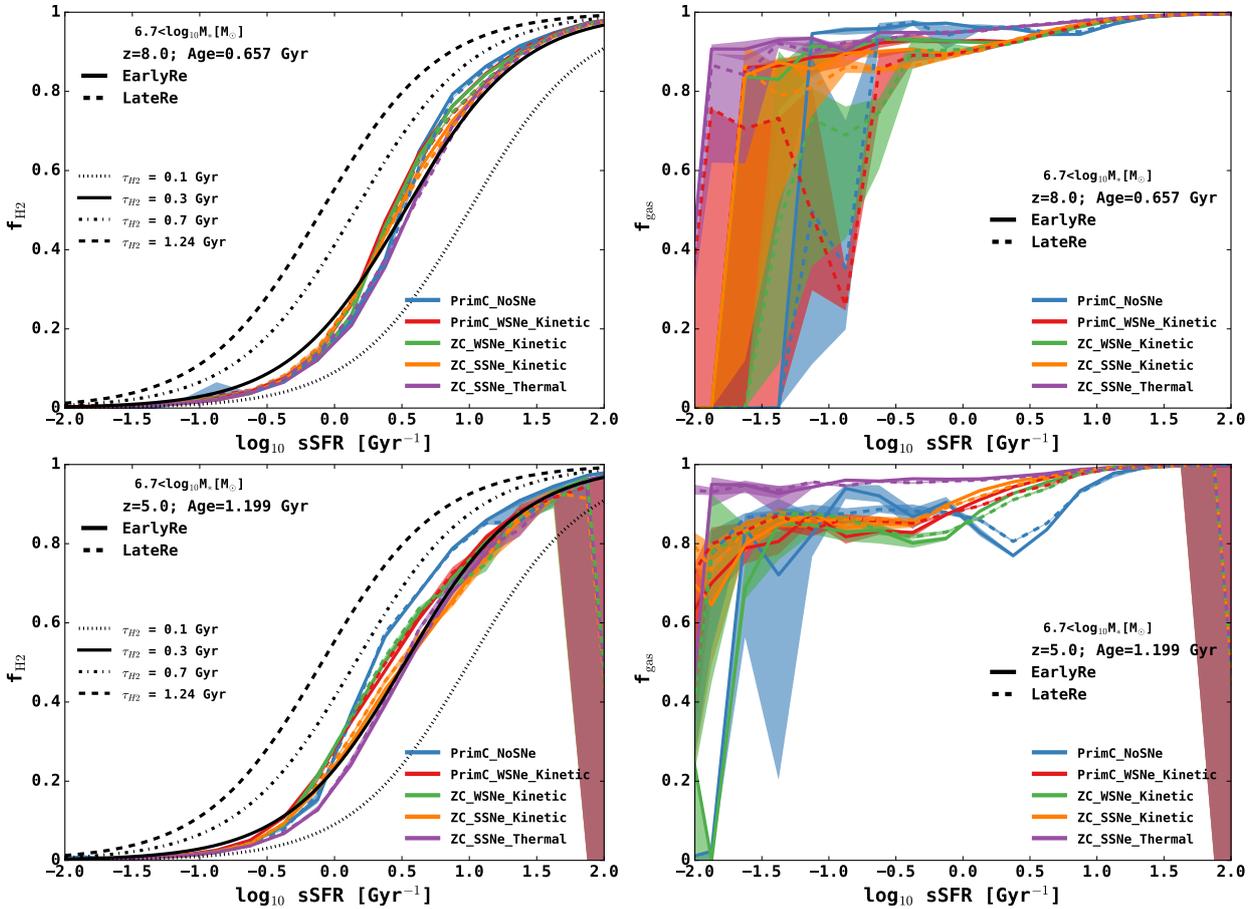

\centering
\begin{subfigure}{\sred\textwidth}
\includegraphics[width=\textwidth,keepaspectratio=true]{{{fh2_vs_ssfr_0512_SMassCut_z7.964357_All}}}
\end{subfigure}\vspace*{-0.25em}
\begin{subfigure}{\sred\textwidth}
\includegraphics[width=\textwidth,keepaspectratio=true]{{{fgas_vs_ssfr_0512_SMassCut_z7.964357_All}}}
\end{subfigure}\vspace*{-0.25em}
\begin{subfigure}{\sred\textwidth}
\includegraphics[width=\textwidth,keepaspectratio=true]{{{fh2_vs_ssfr_0512_SMassCut_z5.000002_All}}}
\end{subfigure}\vspace*{-0.25em}
\begin{subfigure}{\sred\textwidth}
\includegraphics[width=\textwidth,keepaspectratio=true]{{{fgas_vs_ssfr_0512_SMassCut_z5.000002_All}}}
\end{subfigure}\vspace*{-0.7em}
\caption{Here we consider the median fraction of molecular gas in haloes as a function of sSFRs in the left-hand column
for systematically modified galaxy formation models at $z=8$ ($5$) in the top (bottom) panels. In the right-hand column, we consider the total gas fractions
in place of the molecular gas. 
Errorbars are bootstrap-estimated $68$ per cent confidence limits. When a bin has less than three objects we do not denote this with an error-band and instead indicate the value with a dotted-line (this is particularly apparent in the left-hand column at high masses).
For the molecular gas fraction (left column), the dash--dotted line is the~\citet{Tacconi:13} estimate for a constant
  consumption time-scale of $0.7 \gyr$, while the dashed line is the $z=0$ observed time-scale of $1.24 \gyr$~\citep{Saintonge:11a}. 
  All simulations models are better modelled by a faster time-scale of $300 \myr$ (solid line) approximately independent of feedback~\citep{Dave:11a}. We 
  also show a constant time-scale of $100 \myr$, similar to the DRAGONS SAM \textsc{meraxes} at $z=8$ as discussed in Section~\ref{sec:results}
  which is strongly disfavoured.
  For the total gas fraction (right-hand column), there appears to be little trend between the gas fraction and sSFR $\ga 0.1 \gyr^{-1}$.
}\label{fig:f_vs_ssfr}
\end{figure*}

In the left column of Fig.~\ref{fig:f_vs_ssfr} we explore equation (\ref{eqn:fhtwo}) and find that, {\it irrespective of feedback model}, the simulations are approximately fit 
by a constant molecular gas consumption time-scale $\tau_{\rm H2}$ of $300 \myr$ (as we will see in Section~\ref{sec:fhtwo_smass} there is a weak stellar mass dependence). 
We note that this time-scale is shorter than any typically used or estimated for the low-redshift Universe. 
For example,~\citet{Tacconi:13} find $0.7 \gyr$ for main-sequence star-forming galaxies at $z=1-3$ while the same mass systems at $z=0$ consume gas over $1.24\gyr$~\citep{Saintonge:11a}. The analytic arguments of~\citet{Forbes:14a}
result in a cold gas consumption time for present-day haloes that decreases with total mass as $\tau = 3{\rm Gyr} (M_{\rm h} / 10^{12}\Msol)^{-1/2}$~\citep{Forbes:14b},
resulting in consumption time-scales that are several Hubble times for even our most massive objects in \textsc{Smaug} with $M_{\rm h} >10^{11} \Msol$.

\renewcommand{\tabcolsep}{6pt}
\begin{table*}
\caption{The consumption time-scale linking the fraction of molecular hydrogen with the sSFR for different physics schemes,
fit to equation (\ref{eqn:fhtwo}) and shown in the left column of Fig.~\ref{fig:f_vs_ssfr} at key redshifts $z=5 - 8$.}
\label{tab:deplt_fits}
\begin{tabular}{clcccc} \hline Sim & \multicolumn{4}{c}{$\tau_{\rm H2} {\rm [Gyr]}$} \\  & $z=5.0$ & $z=6.0$ & $z=7.0$ & $z=8.0$ \\  \hline PrimC\_NoSNe\_EarlyRe & $0.427\pm{0.004} $  & $0.389\pm{0.005} $  & $0.357\pm{0.002} $  & $0.321\pm{0.002} $  \\ PrimC\_WSNe\_Kinetic\_EarlyRe & $0.361\pm{0.004} $  & $0.384\pm{0.003} $  & $0.376\pm{0.002} $  & $0.345\pm{0.002} $  \\ ZC\_WSNe\_Kinetic\_EarlyRe & $0.367\pm{0.004} $  & $0.374\pm{0.003} $  & $0.368\pm^{0.003}_{0.002} $  & $0.328\pm{0.002} $  \\ ZC\_SSNe\_Kinetic\_EarlyRe & $0.283\pm^{0.005}_{0.004} $  & $0.324\pm{0.004} $  & $0.332\pm{0.003} $  & $0.312\pm{0.002} $  \\ ZC\_SSNe\_Thermal\_EarlyRe & $0.262\pm{0.003} $  & $0.274\pm{0.004} $  & $0.274\pm{0.003} $  & $0.266\pm{0.002} $  \\ PrimC\_NoSNe\_LateRe & $0.418\pm^{0.005}_{0.004} $  & $0.350\pm^{0.001}_{0.002} $  & $0.319\pm{0.002} $  & $0.306\pm{0.002} $  \\ PrimC\_WSNe\_Kinetic\_LateRe & $0.36\pm{0.01} $  & $0.312\pm{0.002} $  & $0.322\pm{0.002} $  & $0.329\pm{0.002} $  \\ ZC\_WSNe\_Kinetic\_LateRe & $0.36\pm{0.01} $  & $0.325\pm{0.002} $  & $0.323\pm{0.002} $  & $0.328\pm{0.002} $  \\ ZC\_SSNe\_Kinetic\_LateRe & $0.30\pm{0.01} $  & $0.272\pm{0.002} $  & $0.291\pm{0.001} $  & $0.308\pm{0.002} $  \\ ZC\_SSNe\_Thermal\_LateRe & $0.28\pm{0.01} $  & $0.247\pm^{0.003}_{0.002} $  & $0.260\pm{0.001} $  & $0.267\pm^{0.002}_{0.001} $  \\  \hline \end{tabular}
\end{table*}
\renewcommand{\tabcolsep}{1pt}

We reiterate that in the left column of Fig.~\ref{fig:f_vs_ssfr}, we see little dependence on stellar feedback ranging from no SNe ($f_{\rm SNe} = 0$) with \emph{PrimC\_NoSNe} in blue to
40 per cent SNe energy ($f_{\rm SNe} = 0.4$) in \emph{PrimC\_WSNe\_Kinetic} in red and ramping up to 100 per cent of the available energy ($f_{\rm SNe} = 1$) in \emph{ZC\_SSNe\_Kinetic} in orange. 
The implementation of the feedback energy, be it kinetic or thermal, appears to have little effect as shown by the similarity between this \emph{ZC\_SSNe\_Kinetic} curve in orange and \emph{ZC\_SSNe\_Thermal} in purple.
This near independence of the consumption time-scale to feedback is in agreement with other, low redshift, studies~\citep[e.g.][]{Dave:11a,Dave:12,Somerville:15}.

We also find little difference between the impact of metal emission-line cooling enhancing the consumption time-scale, as seen by comparing \emph{PrimC\_WSNe\_Kinetic} in red and 
\emph{ZC\_WSNe\_Kinetic} in green. This is unsurprising as it takes time for the metals formed in stars to be released into the halo gas and intergalactic medium.
Additionally, halo virial temperatures must be sufficiently high ($10^6 \rm K$) for heavy metal emission lines to be excited~\citep{Wiersma:09a} and 
thus lead to enhanced cooling of infalling intergalactic medium, or halo, gas.

To explore the best-fitting consumption time-scale in detail, we apply equation (\ref{eqn:fhtwo}) at several redshifts of interest,
using bootstrap sampling of the simulated haloes to estimate confidence intervals. 
As shown in Table~\ref{tab:deplt_fits}, there is a small but significant dependence on the feedback model, with stronger feedback schemes demonstrating progressively {\it shorter} 
molecular consumption time-scales at all redshifts. The apparent contradictory trend of greater feedback resulting in shorter consumption time-scales is explored in Section~\ref{sec:fhtwo_smass}.

The impact of reionization appears to be modest, resulting in consumption time-scales increasing by $30 - 40 \myr$ (or approximately $10-20$ per cent) seen by the close
agreement of the \emph{LateRe} and \emph{EarlyRe} cases by $z=5$. The decrease in SFR as a result of reionization was seen in~\citet{Duffy:14b} for the global SFR density (their Madau diagram, Fig. 6) where the exact redshift of reionization is not so important as the fact that it has occurred. We note that \textsc{Smaug} does not consider self-shielding and hence likely overestimates the (already modest) impact of reionization on this variable.

Overall, we find a remarkable convergence of consumption time-scales of $300 \myr$ for systems spanning the range of possible SNe feedback,
enhanced metallicity cooling and even photoionisation feedback from reionization. This figure is intriguingly close to the gas consumption time-scale estimated by~\citet{Lilly:13} 
based on their `gas regulator' model for galaxies of $M_{\star} = 10^{10} \Msol$ when the Universe is $1 {\rm Gyr}$ old. 
However, we see little evidence in our simulations of their predicted rapid evolution in consumption time-scale (their Fig. 4) in the first billion years of cosmic time.

\subsection{Sensitivity of total gas to sSFR}
The tight and feedback-independent relation between the cold molecular gas phase and the specific star formation in the left column of Fig.~\ref{fig:f_vs_ssfr} can be explored in the context
of the total gas fractions (i.e. all gas gravitationally bound to the subhalo). Similar to equation (\ref{eqn:fhtwo}) we have
\begin{equation}\label{eqn:fgas}
 f_{\rm gas} \equiv M_{\rm gas} / (M_{\rm gas} + M_{\star}) = \frac{1}{1 + (\tau \, \sSFR)^{-1}} \,,
\end{equation}
which we show in the right column of Fig.~\ref{fig:f_vs_ssfr} for a range of redshifts.

Unlike molecular gas, there is no apparent correlation between the total halo gas fraction and sSFR. Instead, we see gas-rich galaxies, with $f_{\rm gas} > 0.8$,
have sSFR that are effectively decoupled from their gas reservoir, exhibiting two to three orders of magnitude variation at all redshifts. This does not appear to fit into the picture of bathtub models, as this total gas fraction is likely dominated by material that lies far from the dense disc material where SF takes place.
As a result, at high-redshift the vast amount of gas in a halo may play little role in setting the local sSFR.
However, the exact transition from gas-poor/ultralow sSFR systems to gas-rich systems (with uncoupled stellar mass doubling times) is feedback dependent. Broadly speaking, a system with strong feedback is likely to transition to a gas-rich system at lower sSFR than a simulation without feedback. This is a result
of that feedback preventing gas from forming stars. Overall, there appears to be a bottleneck in accumulated gas reserves transitioning into a 
useable form for the galactic `economy'. Observations~\citep[e.g.][]{Catinella:10,Saintonge:11a,Saintonge:11b} at low redshift have seen examples of gas rich systems exhibiting 
surprisingly low sSFRs or, equivalently, long consumption time-scales.

Our ability to determine the bottleneck in gas reserves being unable to transition to star-forming material is complicated by the lack of explicit molecular cooling 
in \textsc{Smaug}. However, argued by~\citet{Lagos:15}, who analysed molecular hydrogen in the EAGLE (Evolution and Assembly of GaLaxies and their Environments, \citealt{Schaye:15}) simulations (which share numerous numerical similarities with \textsc{Smaug}), this is a reasonable approach. 
First, the temperature of non-star-forming gas in the surrounding halo is greater than 11000K, 
far above the molecular cooling regime.
Secondly, as found by~\citet{Glover:12}, in the presence of even minor amounts of metals (in particular $C^{+}$), the contribution 
to cooling by $H_2$ is negligible at the high densities for SF. These densities are reached in \textsc{Smaug} and EAGLE alike even though
gas is formally restricted to the atomic phase~\citep{Schaye:04}. Finally, in hydrodynamical simulations such as~\citet{Maio:15}, which do incorporate 
non-equilibrium models for molecular hydrogen production, the impact of $H_2$ is greatest at higher redshifts ($z>10$) than we consider in this work. 

In the next section, we investigate whether the approximate global consumption time-scale is truly uniform across all haloes or, instead, has a more complex dependence on the system it resides in.

\section{Gas consumption dependences}\label{sec:f_smass}
We have shown that the typical consumption time-scale for star-forming gas in the high-redshift Universe is of the order of $300 \myr$,
far shorter than the Hubble time even during the Epoch of Reionization. In detail, however, this time-scale is potentially mass and 
redshift dependent. For example, we might expect smaller stellar mass systems to have lower SFRs and thus a 
much longer consumption time than larger systems, potentially longer than the Hubble time.
Alternatively, at high enough redshift gas inflows may be too fast for SF `demand' to ramp up to meet this supply, 
resulting in consumption times greater than the Hubble time at that epoch. Those systems that have consumption
times less than a Hubble time would be described as supply-side limited, and those that have time-scales greater than the Hubble time
would be demand-side limited. The bathtub / equilibrium models best characterize systems that are supply-side limited (i.e. have adjusted
outflow rates, through SFRs, to the inflow rate). 
Therefore, the suitability of bathtub models can be indicated by the relative consumption time-scale to the Hubble time.

As demonstrated numerically in~\citet{Dave:11b} and discussed in the analytic model of~\citet{Dave:12}, the consumption time-scale for ISM 
gas in haloes is given by
\begin{equation}\label{eqn:dave_deplt}
\tau = 0.4 t_{\rm Hubble} \left( \frac{M_{\star} }{10^{10}\Msol} \right)^{-0.3} \,.
\end{equation}
This result was calculated by assuming a Kennicutt--Schmidt star formation law, in which SFR is set by the gas mass divided by the dynamical time of a disc. The dynamical time-scale
then sets the consumption time-scale which, for a thin disc as given by the~\citealt{MMW:98} model, evolves as the Hubble time, $t_{\rm Hubble}$. Finally, by taking the standard
Kennicutt--Schmidt SF law relating gas surface density with the SF surface density, $\dot{\Sigma}_{\star} \propto \Sigma^{1.4}_{\rm gas}$, and using an empirical relation
between gas surface density and stellar mass, $\Sigma_{\rm gas} \propto M^{3/4}_{\star}$,~\citet{Dave:11b} were able to formulate equation (\ref{eqn:dave_deplt}).
We test whether this scaling holds for the high merger and accretion rate of haloes in the Epoch of Reionization explored by \textsc{Smaug}.

\begin{figure*}
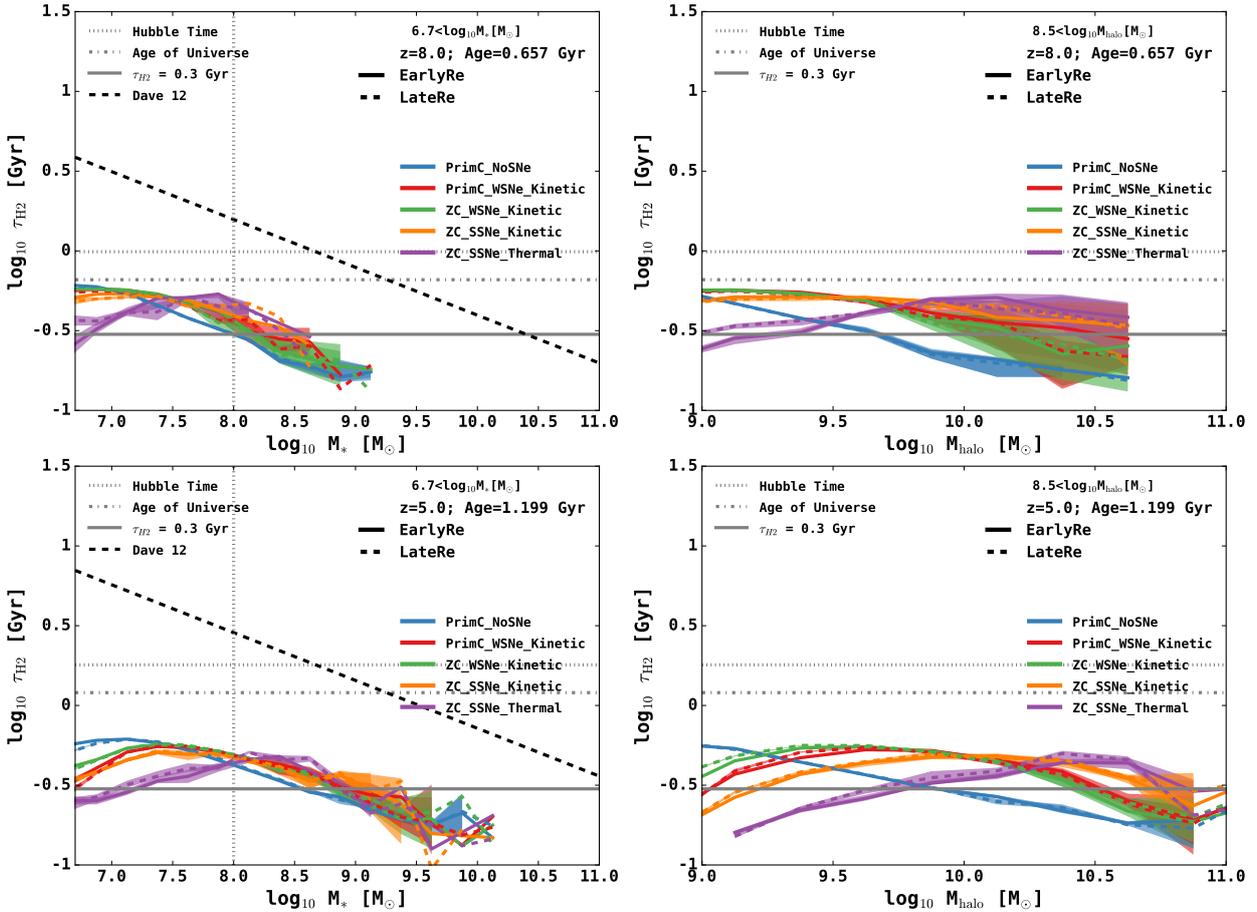

\centering
\begin{subfigure}{\sred\textwidth}
\includegraphics[width=\textwidth,keepaspectratio=true]{{{th2_vs_smass_0512_SMassCut_z7.964357_All}}}
\end{subfigure}\vspace*{-0.2em}
\begin{subfigure}{\sred\textwidth}
\includegraphics[width=\textwidth,keepaspectratio=true]{{{th2_vs_mass_0512_MassCut_z7.964357_All}}}
\end{subfigure}\vspace*{-0.2em}
\begin{subfigure}{\sred\textwidth}
\includegraphics[width=\textwidth,keepaspectratio=true]{{{th2_vs_smass_0512_SMassCut_z5.000002_All}}}
\end{subfigure}\vspace*{-0.2em}
\begin{subfigure}{\sred\textwidth}
\includegraphics[width=\textwidth,keepaspectratio=true]{{{th2_vs_mass_0512_MassCut_z5.000002_All}}}
\end{subfigure}\vspace*{-0.7em}
\caption{Here, we consider the median consumption time, defined as molecular gas mass divided by current SFR 
  as a function of stellar mass in the left-hand column for systematically modified galaxy formation models at $z=8$ ($5$) in the top and bottom
  rows. The right-hand column is the consumption time-scale as a function of halo masses as the same redshifts. 
  Errorbars are bootstrap-estimated $68$ per cent confidence limits. 
When a bin has less than three objects we do not denote this with an error-band and instead indicate the value with a dotted line 
  (this is particularly apparent in the left-hand columns at high masses).
  The horizontal dot--dashed (dotted) black line is the age of the Universe (Hubble time) at the redshift in question. A constant consumption time-scale of $300 \myr$ is shown 
  as a solid horizontal line. The dashed black line is the~\citet{Dave:12} model. The fitting range is for stellar masses greater than the vertical dashed line, with best
  fits shown in Table~\ref{tab:th2_smass_fits}. }\label{fig:th2_deplt}
\end{figure*}

\begin{figure*}
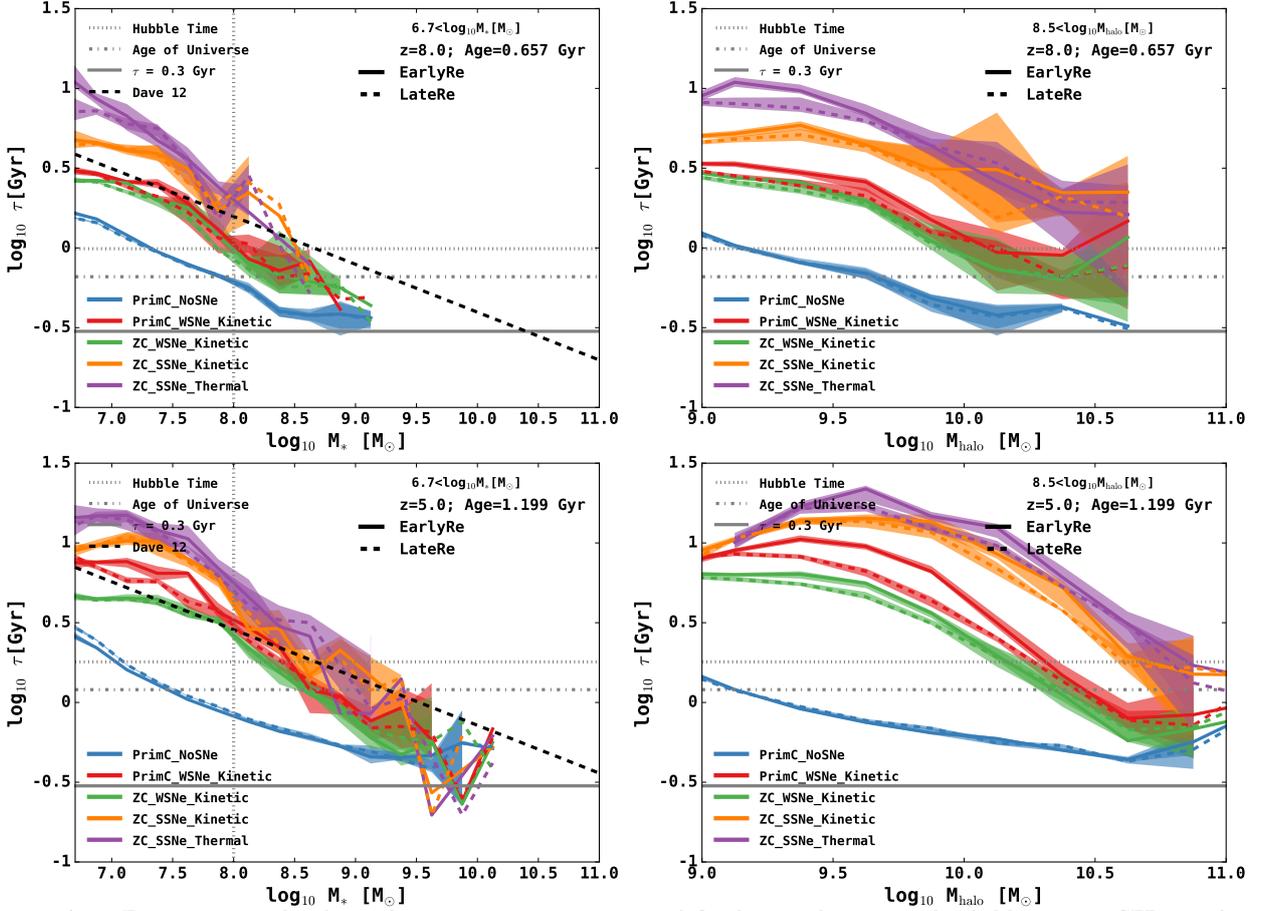

\centering
\begin{subfigure}{\sred\textwidth}
\includegraphics[width=\textwidth,keepaspectratio=true]{{{tau_vs_smass_0512_SMassCut_z7.964357_All}}}
\end{subfigure}\vspace*{-0.2em}
\begin{subfigure}{\sred\textwidth}
\includegraphics[width=\textwidth,keepaspectratio=true]{{{tau_vs_mass_0512_MassCut_z7.964357_All}}}
\end{subfigure}\vspace*{-0.2em}
\begin{subfigure}{\sred\textwidth}
\includegraphics[width=\textwidth,keepaspectratio=true]{{{tau_vs_smass_0512_SMassCut_z5.000002_All}}}
\end{subfigure}\vspace*{-0.2em}
\begin{subfigure}{\sred\textwidth}
\includegraphics[width=\textwidth,keepaspectratio=true]{{{tau_vs_mass_0512_MassCut_z5.000002_All}}}
\end{subfigure}\vspace*{-0.7em}
\caption{As in Fig.~\ref{fig:th2_deplt}, we consider the median consumption time, now defined as total gas mass divided by current SFR
  as a function of stellar mass in the left column for systematically modified galaxy formation models at $z=8$ ($5$) in the top and bottom rows. 
  The right column is the same consumption time-scale as a function of halo mass for those redshifts. 
  Errorbars are bootstrap-estimated $68$ per cent confidence limits. 
  The horizontal dot--dashed (dotted) grey line is the age of the Universe (Hubble time) at the redshift in question. A constant consumption time-scale of $300 \myr$ is shown 
  as a horizontal solid line. The fitting range is for stellar masses greater than the vertical dashed line, with best
  fits shown in Table~\ref{tab:deplt_smass_fits}. }\label{fig:tau_deplt}
\end{figure*}

\subsection{Cold molecular gas}\label{sec:fhtwo_smass}
In the left column of Fig.~\ref{fig:th2_deplt}, we explore how the typical molecular gas consumption time-scale
$\tau_{\rm H2}$ varies as a function of total stellar mass across a range of redshifts. We denote equation (\ref{eqn:dave_deplt})
with a black dashed curve and the Hubble time (age of the Universe) with dotted (dot--dashed) grey horizontal lines at the redshift in question.
The fiducial $300\myr$ time is indicated by the grey solid horizontal line.

In the left column of Fig.~\ref{fig:th2_deplt}, it is clear that, at all stellar masses and redshifts considered, 
the consumption time-scale $\tau_{\rm H2}$ is less than the Hubble time. These time-scales are at least an order of magnitude faster than equation (\ref{eqn:dave_deplt})
suggests, although the total gas is in much better agreement (left column of Fig.~\ref{fig:tau_deplt}). This is somewhat surprising as the molecular gas fraction is created from the 
equation of state, i.e. ISM, gas in \textsc{Smaug} and hence is closely tied to the `gas' term adopted in~\citet{Dave:12}. We will return to this issue in the next section. 

To explore the relative insensitivity of the molecular consumption time-scale to feedback more quantitatively, we fit the evolving power law of equation (\ref{eqn:evopowerlaw})
over the redshift range of $5-10$, with best fits (and bootstrap-estimated errors) shown in Table~\ref{tab:th2_smass_fits}.
Overall for a system of $M_{*} = 10^{8} \hMsol$ at $z=7$, the molecular gas is depleted in just $400 \myr$. 

As shown in Table~\ref{tab:th2_smass_fits} there is a modest dependence on stellar mass, with $B_{\rm mass} \approx -0.3$ for all simulations (in agreement with~\citealt{Dave:12}). 
As a result of this sublinear dependence, consumption time-scales change by only factors of 2-3 across orders of magnitude in stellar mass. 
Due to the relatively limited population of large stellar mass systems in strong feedback models, the errors on their best-fitting solutions are rather large. Therefore, when we explore
the use of a varying depletion time-scale (Section~\ref{sec:sam_tau}), we select the robustly estimated fits for \emph{ZC\_WSNe\_Kinetic\_EarlyRe}; however, all simulations have
approximately the same qualitative behaviour.

Finally, we note that there appears to be a turnover at masses below our minimum fitting mass \mstarmin (particularly apparent for the strong feedback \emph{ZC\_SSNe\_Thermal} 
model). We denote this as a dashed vertical line in the left column of Fig.~\ref{fig:th2_deplt} for clarity and note that choice of this cut-off is critical for the resulting power law behaviour.
In Fig.~\ref{fig:restest_tau}, we explore resolution tests that show that this turnover is not apparently a simple consequence of numerical resolution. 
We leave this low-mass behaviour for zoom-in simulations to be presented in future work but note that the low-mass systems tend to have far less molecular gas available making the
depletion time-scale of this gas relatively unimportant.

Overall, the consumption time-scale appears to have only a modest evolution across redshift ($ -1.2 < C_{\rm evo} < -0.9$) 
that was already suggested by the approximately constant consumption time-scale shown in Table~\ref{tab:deplt_fits}. 
Overall, this is significantly less rapid evolution than the model of~\citet{Dave:12} would suggest, which scales as $t_{\rm H}$ [and hence $(1+z)^{-1.5}$ in the matter era], 
but is in much greater agreement with the empirically determined time-scale evolution of~\citet{Genzel:15}.

In the right column of Fig.~\ref{fig:th2_deplt} we consider the consumption time-scales as a function of halo mass that now show greater variation as a function of feedback. 
The stellar mass independence of the various feedback schemes in the left column is revealed to be a result of a changing stellar-to-halo mass relation. 
This now explains the counterintuitive {\it average} consumption time in Section~\ref{sec:gas_frac} in which strong feedback schemes
more rapidly consume molecular gas reserves. The strong feedback systems have shorter consumption time-scales for low halo mass systems (as seen in the left column 
of Fig.~\ref{fig:th2_deplt}) that dominate the sample size and hence global consumption time-scale for Fig.~\ref{fig:f_vs_ssfr} and Table~\ref{tab:deplt_fits}.

The overall picture of galaxies growing at high-redshift is one of rapid consumption time-scales that are insensitive to the feedback scheme used but not the stellar mass of the system
they find themselves in. The situation of the gas reservoir surrounding the early galaxies is far more sensitive to the modelled feedback, as we explore in the next section.
 
\begin{table*}
\caption{The relation between the molecular consumption time-scale considered as a function of the stellar mass fit to equation (\ref{eqn:evopowerlaw})
given by $\tau_{\rm H2} = 10^{A_{\tau_{\rm H2}}} \left[{M_{\star}}/{ M_{\rm pivot}} \right]^{B_{\rm mass}}\left[ (1+z)/(1+z_{\rm pivot}) \right]^{C_{\rm evo}}$
for objects with mass greater than \mstarmin and over the redshift range of $5-10$. The pivot mass is \mstarpivot and pivot redshift is 7. 
There is little difference between the various schemes although the strong feedback models have larger errors due to the fewer high stellar mass systems they form.}
\label{tab:th2_smass_fits}
\begin{tabular}{clccc} \hline Sim & $A_{\tau_{H2}} {\rm [Gyr]}$ & $B_{\rm mass}$ & $C_{evo}$ \\ \hline PrimC\_NoSNe\_EarlyRe & $-0.75\pm{0.03} $  & $-0.25\pm{0.02} $  & $-1.06\pm^{0.10}_{0.13} $  \\ PrimC\_WSNe\_Kinetic\_EarlyRe & $-0.69\pm^{0.04}_{0.07} $  & $-0.28\pm^{0.01}_{0.05} $  & $-1.19\pm^{0.12}_{0.11} $  \\ ZC\_WSNe\_Kinetic\_EarlyRe & $-0.71\pm^{0.04}_{0.08} $  & $-0.29\pm^{0.02}_{0.06} $  & $-1.13\pm^{0.13}_{0.14} $  \\ ZC\_SSNe\_Kinetic\_EarlyRe & $-0.66\pm^{0.03}_{0.09} $  & $-0.31\pm^{0.02}_{0.05} $  & $-1.01\pm{0.09} $  \\ ZC\_SSNe\_Thermal\_EarlyRe & $-0.67\pm^{0.03}_{0.07} $  & $-0.32\pm{0.02} $  & $-0.93\pm^{0.11}_{0.24} $  \\ PrimC\_NoSNe\_LateRe & $-0.74\pm^{0.03}_{0.04} $  & $-0.23\pm^{0.02}_{0.04} $  & $-1.09\pm^{0.11}_{0.10} $  \\ PrimC\_WSNe\_Kinetic\_LateRe & $-0.69\pm^{0.05}_{0.08} $  & $-0.28\pm^{0.02}_{0.07} $  & $-1.06\pm^{0.09}_{0.11} $  \\ ZC\_WSNe\_Kinetic\_LateRe & $-0.72\pm^{0.04}_{0.06} $  & $-0.30\pm^{0.02}_{0.04} $  & $-1.24\pm^{0.10}_{0.06} $  \\ ZC\_SSNe\_Kinetic\_LateRe & $-0.69\pm^{0.02}_{0.03} $  & $-0.34\pm^{0.03}_{0.02} $  & $-1.06\pm{0.06} $  \\ ZC\_SSNe\_Thermal\_LateRe & $-0.67\pm^{0.03}_{0.06} $  & $-0.34\pm^{0.01}_{0.02} $  & $-0.93\pm^{0.11}_{0.10} $  \\  \hline \end{tabular}
\end{table*}

\begin{table*}
\caption{The relation between the gas consumption time-scale considered as a function of the stellar mass fit to equation (\ref{eqn:evopowerlaw}) given by 
$\tau = 10^{A_{\tau}} \left[{M_{\star}}/{ M_{\rm pivot}} \right]^{B_{\rm mass}}\left[ (1+z)/(1+z_{\rm pivot}) \right]^{C_{\rm evo}}$
for objects with mass greater than \mstarmin (shown as a vertical dashed line in the relevant plots) and over the redshift range of $5-10$. 
The pivot mass is \mstarpivot and pivot redshift is 7. Increasing feedback greatly increases total gas consumption time-scales.}
\label{tab:deplt_smass_fits}
\begin{tabular}{clccc} \hline Sim & $A_{\tau} {\rm [Gyr]}$ & $B_{\rm mass}$ & $C_{evo}$ \\ \hline PrimC\_NoSNe\_EarlyRe & $-0.40\pm^{0.04}_{0.07} $  & $-0.16\pm^{0.03}_{0.07} $  & $-1.03\pm^{0.10}_{0.18} $  \\ PrimC\_WSNe\_Kinetic\_EarlyRe & $-0.23\pm^{0.06}_{0.13} $  & $-0.38\pm^{0.03}_{0.10} $  & $-1.96\pm^{0.16}_{0.19} $  \\ ZC\_WSNe\_Kinetic\_EarlyRe & $-0.27\pm^{0.06}_{0.12} $  & $-0.37\pm^{0.02}_{0.06} $  & $-1.62\pm^{0.15}_{0.31} $  \\ ZC\_SSNe\_Kinetic\_EarlyRe & $-0.14\pm^{0.05}_{0.15} $  & $-0.54\pm{0.03} $  & $-2.20\pm^{0.15}_{0.44} $  \\ ZC\_SSNe\_Thermal\_EarlyRe & $-0.19\pm^{0.04}_{0.17} $  & $-0.64\pm^{0.07}_{0.06} $  & $-2.75\pm^{0.23}_{0.28} $  \\ PrimC\_NoSNe\_LateRe & $-0.40\pm^{0.04}_{0.07} $  & $-0.17\pm^{0.03}_{0.07} $  & $-1.11\pm^{0.11}_{0.20} $  \\ PrimC\_WSNe\_Kinetic\_LateRe & $-0.23\pm^{0.06}_{0.12} $  & $-0.36\pm^{0.02}_{0.08} $  & $-1.70\pm^{0.14}_{0.32} $  \\ ZC\_WSNe\_Kinetic\_LateRe & $-0.28\pm^{0.06}_{0.10} $  & $-0.38\pm^{0.03}_{0.05} $  & $-1.83\pm^{0.12}_{0.17} $  \\ ZC\_SSNe\_Kinetic\_LateRe & $-0.16\pm^{0.04}_{0.11} $  & $-0.57\pm{0.04} $  & $-2.16\pm^{0.15}_{0.30} $  \\ ZC\_SSNe\_Thermal\_LateRe & $-0.19\pm^{0.04}_{0.13} $  & $-0.67\pm^{0.04}_{0.03} $  & $-2.72\pm^{0.27}_{0.38} $  \\  \hline \end{tabular}
\end{table*}

\subsection{Broader supply chains}
We now consider the entire gas consumption time-scale in haloes, as given in Fig.~\ref{fig:tau_deplt}, which is far longer than the corresponding molecular consumption time.
We see that without feedback, \emph{PrimC\_NoSNe}, galaxies of $\log_{10} M_{*} > 7.5$ have gas consumption time-scales less than a Hubble time at all redshifts considered. 
No-feedback models exhibit `run-away' SF and are effectively supply-side limited. This means that a model without feedback effectively reaches equilibrium in the 
Epoch of Reionization for systems with $\log_{10} M_{*} > 7.5$ such that the SFR traces the inflow accretion rate.

However, {\it any} feedback then moves the system to a quasi-regulated scenario in which the gas consumption time-scales are all larger than a Hubble time
for $\log_{10} M_{*} < 8.5$, which in economics parlance is termed demand-side limited. For masses greater than this, we return to a supply-side limited scenario.
We therefore conclude that the basic assumption of the bathtub model in which galaxies can adjust total gas consumption within a Hubble time does not apply for the dwarf galaxies 
responsible for providing most of the UV photons to reionize the high-redshift Universe~\citep{Duffy:14b}. 

Although the normalization of the gas consumption time-scale is strongly feedback dependent, the stellar-mass-dependent scaling argued by~\citet{Dave:12} appears to be
a good description. This is actually surprising because the gas considered in our analysis is the entire gas distribution in the halo, and not the molecular hydrogen component
traced by the~\citet{Dave:12} model. As before, we fit the evolving power law of equation (\ref{eqn:evopowerlaw}) to our simulations 
over the redshift range of $5-10$, with best fits (and bootstrap-estimated errors) shown in Table~\ref{tab:deplt_smass_fits}.

For the no-feedback case we obtain a more modest mass dependence ($B_{\rm mass} \sim -0.16$ compared to $-0.3$) to that found by~\citet{Dave:12}. 
For models with feedback there is a modest steepening of the relation from $B_{\rm mass} \approx -0.4$ to $-0.7$ as a function of increasing feedback. The overall normalization
from the~\citet{Dave:12} model agrees well with the kinetic SNe feedback scheme \emph{ZC\_WSNe\_Kinetic}, which is most similar to their simulation,
as detailed in~\citet{Dave:11a, Dave:11b}.

Of all the physics schemes in \textsc{Smaug} the one that best matches the observed SFR function at high-redshift is \emph{ZC\_SSNe\_Thermal}~\citep{Duffy:14b} which has over twice the SNe energy available for feedback than both \emph{ZC\_WSNe\_Kinetic} and the model of~\citet{Dave:12}. As shown in the left column of Fig.~\ref{fig:tau_deplt} in purple, 
this model also has over twice the gas consumption time-scale. This suggests that we are able to reproduce high-redshift observations by increasing total gas consumption time-scales
through SNe feedback by disrupting and likely ejecting the halo gas that would otherwise transition into the cold star-forming phase.

The variation in the best-fitting evolutions of the consumption time-scales, $C_{\rm evo}$, is a key sign that removal (and recycling from winds, e.g.,~\citealt{Oppenheimer:10}) of 
material can significantly impact the simple scalings presented in bathtub models of galaxy formation. Overall, however, models with feedback have 
$C_{\rm evo}\approx -2$ to $-3$ which is a far stronger redshift dependence than the no-feedback case, $C_{\rm evo} \approx -1$. 
Between these two extremes lies the scaling from~\citet{Dave:12} which argued that the evolution is linear in Hubble time, 
i.e. $C_{\rm evo}=-1.5$ as $t_{\rm Hubble} \propto (1+z)^{-1.5}$ in the matter-dominated era at high-redshift. 

The order of magnitude difference between the molecular gas and total gas consumption time-scales again demonstrates that there exists a bottleneck in high-redshift galaxies in
converting infalling gas into the cold, dense star-forming phase~\citep[as also argued by][]{Krumholz:12b}. 
Broadly speaking, our {\it total} gas consumption time-scales at $z>5$ are similar to those observed in much larger starburst systems at lower redshift, $z=1-2$, with 
$\tau \approx 0.3-2$ \gyr~\citep{Daddi:10,Tacconi:10,Tacconi:13}. This suggests that while galaxies during the Epoch of Reionization have similarly rapid 
consumption of gas to low-redshift massive systems, the reason is different with merger-induced starbursts replicating the ubiquitous inflow fed SFRs of the high-redshift Universe.

\begin{figure*}
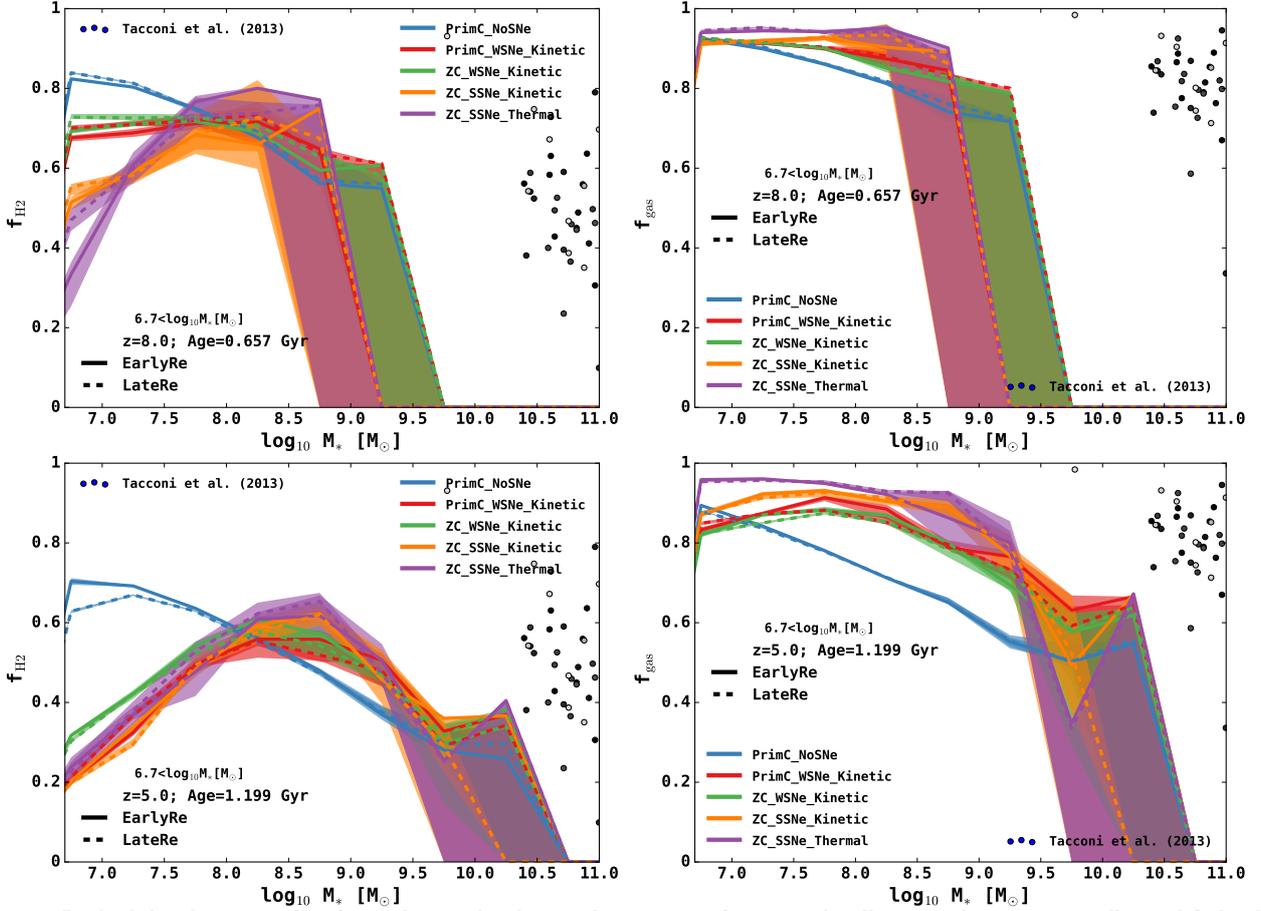

\centering
\begin{subfigure}{\sred\textwidth}
\includegraphics[width=\textwidth,keepaspectratio=true]{{{fh2_vs_smass_0512_SMassCut_z7.964357_All}}}
\end{subfigure}\vspace*{-0.2em}
\begin{subfigure}{\sred\textwidth}
\includegraphics[width=\textwidth,keepaspectratio=true]{{{fgas_vs_smass_0512_SMassCut_z7.964357_All}}}
\end{subfigure}\vspace*{-0.2em}
\begin{subfigure}{\sred\textwidth}
\includegraphics[width=\textwidth,keepaspectratio=true]{{{fh2_vs_smass_0512_SMassCut_z5.000002_All}}}
\end{subfigure}\vspace*{-0.2em}
\begin{subfigure}{\sred\textwidth}
\includegraphics[width=\textwidth,keepaspectratio=true]{{{fgas_vs_smass_0512_SMassCut_z5.000002_All}}}
\end{subfigure}\vspace*{-0.7em}
\caption{In the left column we plot the median molecular gas fraction as a function of stellar mass for systematically modified galaxy 
formation models at two redshifts of interest, $z=8$ and $5$, in the top and bottom rows respectively. We also explore the total gas
fractions as a function of stellar mass in the right column for the same redshifts. Errorbars are bootstrap-estimated $68$ per cent confidence limits. 
The data points are the total or molecular gas fractions for low redshift samples from~\citet{Tacconi:13} for redshifts from $z=1$ to 
$z=2.3$ with higher redshifts indicated with darker shading.
}\label{fig:f_vs_smass}
\end{figure*}

\subsection{The 300 million year question}
We now consider possible physical causes for the rapid and consistent molecular consumption time-scale of $300 \myr$, as well as linear dependence on redshift. We emphasize that in the simulations presented in this study and in the following analysis we consider only a constant density threshold for SF of the form given in~\citet{Schaye:08}. 

We take the ratio of the molecular gas (calculated using the empirical scaling with local pressure of~\citealt{Leroy:08}) as defined in equation (\ref{eqn:molmass})
and the SFR, which is given by the pressure-based Kennicutt--Schmidt law~\citep{Kennicutt:98b} derived by~\citet{Schaye:08}:
\begin{equation}\label{eqn:sfrlaw}
 \dot{M}_{\star} =  M_{\rm gas} A (1 M_{\odot} {\rm pc}^{-2})^{-n} \left(\frac{\gamma}{G} f_{\rm gas} P \right)^{(n-1)/2}\,,
\end{equation}
with ratio of specific heats $\gamma=5/3$, the gas fraction $f_{\rm gas}$ that is effectively unity for our systems, 
$A=1.515 \times 10^{-4} M_{\odot} {\rm yr^{-1} kpc^{-2}} $ and $n=1.4$ set by the observed~\citet{Kennicutt:98b} relation. 
The pressure for dense (i.e. above ${n}_{\rm crit} = 0.1 \,{\rm cm^{-3}}$) star-forming gas is governed by a polytropic equation of state as our simulations lack the resolution to track
this otherwise multiphase medium
\begin{equation}\label{eqn:pressurelaw}
P = P_{\rm crit}\left( \frac{n_{\rm H}}{{n}_{\rm crit}} \right)^{\gamma_{\rm eff}}\,,
\end{equation}
where $P_{\rm crit}/k = 1.08 \times 10^{3}\,{\rm K}\,{\rm cm}^{-3}$, $\gamma_{\rm eff}=4/3$. The resultant molecular consumption time $\tau_{\rm H2}$ for a single gas particle is then the ratio of equations (\ref{eqn:molmass}) and (\ref{eqn:sfrlaw}), giving
\begin{equation}\label{eqn:deplttime_first}
\tau_{\rm H2} = \frac{(1 M_{\odot} {\rm pc}^{-2})^{n}}{A(1+R_{\rm surf}^{-1})}  \left(\frac{\gamma}{G} f_{\rm gas} P \right)^{(1-n)/2}\,.
\end{equation}
We can then determine that a consumption time-scale of $300 \myr$ corresponds to gas with particle density $\mathcal{O}(500 \, {\rm cm^{-3}})$, which is an extremely high number density for galaxies
at low redshift but similar to giant molecular clouds in their self-bound collapse phase~\citep{Krumholz:12a}. The SF in the central regions of the galaxies at high-redshift is effectively 
an enormous molecular cloud, with over a third of the mass in the most massive systems above this density at $z=5$. We note that the median logarithmic density across all star-forming particles is of the order of $10\, {\rm cm^{-3}}$, which corresponds to consumption time-scales only twice that of the highest density gas.
That fact that feedback can make little difference to the molecular gas consumption time-scale is reasonable given the extraordinarily high density of the gas contributing to this phase.

We can also explore the redshift dependence of the consumption time-scale. We start by using the~\citet{Schaye:04} calculation for 
the central particle number density $n_{H}$ of a self-gravitating disc assuming an exponential surface profile, 
\begin{equation}\label{eqn:particlerho}
n_{\rm H} = \frac{G M^{2}_{\rm disc}}{12 {\upi} c_{\rm s}^{2} R^{4}_{\rm disc} \mu m_{\rm p}}\,,
\end{equation}
where the mass of the disc $M_{\rm disc} = 0.17 M_{\rm halo}$ with characteristic disc scalelength $R_{\rm disc} = \lambda R_{\rm halo} / \sqrt{2}$,
with spin parameter $\lambda \approx 0.05$ at high-redshift~\citep{Angel:16}. We further assume the sound speed $c_{\rm s}=10 \kms$ for an isothermal gas at $10^4 {\rm K}$,
$\mu = 1.22$ as the molecular weight for primordial neutral gas and $m_{\rm p}$ is the mass of hydrogen. Using the stellar-to-halo mass relation ($M_{\star}  = 10^{6.5} \Msol (M_{\rm halo} / 10^{9} \Msol)^{7/5}$)
from~\citet{Mutch:16a} we then can show that $z=5$ haloes of stellar mass $10^{9}\Msol$ can be expected to have central densities of the order of $500 \, {\rm cm^{-3}}$,
which is exactly what the \textsc{Smaug} simulations show, and that a consumption time-scale of $300 \myr$ requires. 

From equation (\ref{eqn:particlerho}), we can see that the number density $n_{\rm H}$ naturally scales as $(1+z)^4$ assuming constant fraction of disc length to virial radius ($R_{\rm halo}$).
We can then expect that our star-forming gas pressure, equation (\ref{eqn:pressurelaw}), in the disc will scale as $(1+z)^{4\gamma_{\rm eff}}$. This pressure-redshift scaling then 
leads to a molecular gas consumption time-scale (equation \ref{eqn:deplttime_first}) that depends on redshift as 
$\tau_{\rm H2}  \propto (1+z)^{4\gamma_{\rm eff}(n-1)/2} \approx (1+z)^{-1.06}$. This result is in excellent agreement with our fits in Table~\ref{tab:th2_smass_fits} 
and our adopted case of \emph{ZC\_WSNe\_Kinetic\_EarlyRe} with $\tau_{\rm H2} = 190 {\rm Myr} \left[{M_{\star}}/{ 10^{9} \Msol } \right]^{-0.3}\left[ (1+z)/8 \right]^{-1.1}$.
Overall, the hydrodynamical simulations suggest a varying consumption time-scale, we will explore the incorporation
of this time-scale in SAMs of the Epoch of Reionization in Section~\ref{sec:sam_tau}.

\section{Gas distribution}\label{sec:gas_dist}
With the increasing sensitivity of mm-wavelength facilities, such as ALMA and IRAM, there are now statistical samples of the gas contents of galaxies
at moderately high-redshifts, in particular we will compare our simulations to \citet{Tacconi:13} for $z<2.4$ galaxies. A key result from
bathtub models is that high-redshift galaxies will be gas rich, and only in the most massive systems can the gas reserves be depleted faster than they are replenished
by infall (a prediction that we confirmed in the last section).

In the left column of Fig.~\ref{fig:f_vs_smass}, we show the molecular gas fraction in haloes as a function of stellar mass for several high-redshift examples. Overall, systems 
that exhibit run-away SF as they have no stellar feedback (\emph{PrimC\_NoSNe} in blue) are the most gas rich at low stellar masses. This is the same
result we saw for the consumption time-scale in the left column of Fig.~\ref{fig:th2_deplt}, and shows that at low masses SFRs simply cannot
respond to the rapid inflow of material. Therefore, gas-rich systems are demand-side limited. The effect of increasing stellar feedback is that gas is removed in lower
mass systems. Above $\log_{10} M_{*} > 8$, all simulations show the same decrease of gas fraction with increasing stellar mass, with $z=5$ systems $20-30$ per cent less gas rich than at $z=8$.

From~\citet{Dave:11b}, the expected scaling at low redshifts is approximately $f_{\rm gas} \propto M^{-0.5}_{\star}$, which is far steeper than we see at high 
redshift, confirming the picture that high-redshift systems are more gas rich than low-redshift counterparts. Indeed, it has been argued that gas-poor systems at late times are a 
natural consequence of the bathtub model because the accretion rate drops more rapidly with redshift than the SFR, implying that the gas reservoir is slowly used up.

Intriguingly, the high molecular fractions from~\citet{Tacconi:13} are only reproduced in our high-redshift objects with an order of magnitude lower stellar mass. 
To match their observations would require the characteristic stellar mass scale at which molecular gas fractions peak to increase with decreasing redshift.
Thus, while it may be possible to match the high gas fraction results in the lower redshift observations of~\citet{Tacconi:13} by a secular evolution in the galaxy populations,
it is unclear what would drive this trend. One possible mechanism could be mergers bringing in significant new material, as well as increasing the pressure such that the molecular to atomic gas ratio increases. 
Unfortunately, we cannot test this in \textsc{Smaug} as the simulation does not continue below $z=5$.

Although there is significant scatter in the gas fraction with feedback models, we see a modest trend that {\it stronger} SNe typically result in more gas-rich systems. 
This is a consequence of increased feedback energy requiring fewer exploding stars to suppress SF in the surrounding gas. In other words, a given mass of stars formed in a strong feedback model will eject more gas from the diffuse halo reservoir, preventing that gas being locked up in stars and 
increasing the gas fraction. We note that as the gas fraction is defined as gas mass divided by the total gas and stellar masses, a modest reduction in gas mass is less significant in reducing the gas fraction than allowing (even some of) that ejected gas to form stars. 
It is entirely possible for some of this ejected gas mass to be later re-accreted billions of years later as a wind-mode recycling of material (e.g.~\citealt{Oppenheimer:10}) but not enough time has elapsed by the end of the Epoch of Reionization for this to occur.

In conclusion, the simplest interpretation of the difference in consumption time-scales of molecular gas (Fig.~\ref{fig:th2_deplt}) 
and total gas (Fig.~\ref{fig:tau_deplt}) at fixed halo mass is that the amount of star-forming gas (from which the molecular gas is entirely sourced) 
is suppressed relative to the halo gas mass. The challenge with this picture is that feedback is expected to remove lower density material preferentially to the dense, star-forming gas. We are therefore left with the conclusion that feedback has acted to restrict the inflow of gas into the dense phase, either through
direct heating of star-forming gas, or more likely disruption and possibly ejection of the halo material transitioning to that phase.

\begin{figure*}
\centering
\begin{subfigure}{\sred\textwidth}
\includegraphics[width=\textwidth,keepaspectratio=true]{{{SMF_Meraxes_z11.000000_All}}}
\end{subfigure}\vspace*{-0.2em}
\begin{subfigure}{\sred\textwidth}
\includegraphics[width=\textwidth,keepaspectratio=true]{{{SMF_Meraxes_z6.986987_All}}}
\end{subfigure}\vspace*{-0.2em}
\begin{subfigure}{\sred\textwidth}
\includegraphics[width=\textwidth,keepaspectratio=true]{{{SMF_Meraxes_z5.000002_All}}}
\end{subfigure}\vspace*{-0.2em}
\begin{subfigure}{\sred\textwidth}
\includegraphics[width=\textwidth,keepaspectratio=true]{{{SMF_Meraxes_z6.006774_All}}}
\end{subfigure}\vspace*{-0.7em}
\caption{Here we consider the predicted stellar mass functions at $z=11,\, 7,\, 6,\, 5$ (top left clockwise) from the \textsc{meraxes} semi-analytic model with the
default star formation law from~\citet{Mutch:16a} as `Meraxes' in red. We then compare this with a modified \textsc{meraxes} model using the a star formation 
law with a constant molecular consumption time-scale, a scheme based on~\citet{Croton:06}. The blue line denotes $\tau_{\rm H2} \approx 2 {\rm Gyr}$ set by 
local Universe measurements~\citep{Leroy:08}. We then explore $300 \myr$ (in green) as suggested by \textsc{Smaug} for high-redshifts
and a variable time-scale in purple dependent on mass and redshift as given by Table~\ref{tab:th2_smass_fits} and explained in the text. Overall
the reduced time-scales agree better with the observations of~\citet{Song:16} than the long duration time-scale. 
We note that we require the redshift evolution plus short consumption time-scale to produce systems as large as `GN-z11'~\citep{Oesch:16} at $z=11$, as~\citet{Mutch:16b} 
achieved in \textsc{meraxes}. However, we find that a shorter time-scale is still unable to explain the observed downturn in stellar mass $<10^{9} \Msol$ at $z=6\, \&\, 7$~\citep{Gonzalez:11,Duncan:14} as does the original \textsc{meraxes} formalism of~\citet{Mutch:16a}.}\label{fig:deplt_sam}
\end{figure*}

\section{Discussion}\label{sec:discussion}\label{sec:results}
Before discussing how our results might be used to improve the SF recipes in SAMs for galaxy formation at high-redshift we first summarize our main findings.

We have found that \textsc{Smaug} hydrodynamic simulations predict the molecular gas fraction to strongly correlate with the sSFR, 
with a typical molecular consumption time-scale of $300 \myr$ (as given in the left column of Fig.~\ref{fig:f_vs_ssfr}) irrespective of feedback. 
In detail this time-scale has a modest dependence on feedback, with shorter consumption time-scales for increasingly
strong SNe feedback, as well as decreasing redshift, as shown in Table~\ref{tab:deplt_fits}. Typically, reionization acts to increase the consumption time-scale by
10-20 per cent ($30-60 \rm Myr$). We found no such trends with the total gas fraction. For values above ultralow sSFR (with doubling times an order
of magnitude longer than the Hubble time), all objects were similarly gas rich ($f_{\rm gas} > 0.8$) irrespective of their growth rate.

We then explored in greater detail the dependences of the consumption time-scales. 
As shown in Fig.~\ref{fig:th2_deplt} (and with best fits in Table~\ref{tab:th2_smass_fits}), the molecular gas consumption time-scale depends on stellar mass and redshift. 
However, we find that at all masses explored the star-forming (molecular) gas phase is consumed far more quickly than a Hubble time 
and in a manner independent of the SNe feedback physics. 
As a result, SF in these systems is constrained by how quickly infalling gas is converted into the molecular phase. 

The situation is greatly changed when we consider the total gas consumption time-scales, shown in Fig.~\ref{fig:tau_deplt} and best fits in Table~\ref{tab:deplt_smass_fits}, which 
is strongly dependent on the feedback model. We find that consumption times increase at fixed stellar mass and redshift as a function of increasing feedback energy. 
This is distinct from bathtub model predictions (which to first order are feedback-independent time-scales) and is a critical
reminder that the amount of gas is not so important as the state of that gas, even in a cold and dense early universe.

\subsection{Implications for semi-analytic galaxy formation models}\label{sec:sam_tau}
Implementation of findings from this investigation in SAMs of the Epoch of Reionization depend 
on the particular SF parameterisation. In this section, we consider the SAMs presented in~\citet[][\textsc{meraxes}]{Mutch:16a} and~\citet[][\textsc{galform}]{Lagos:11a}.

The original \textsc{galform} model that adopted $ \dot{M}_{\rm \star} =  {M}_{\rm cold}  / \tau_{\rm \star}$~\citep{Cole:00} could be readily adaptable to the simple constant consumption time-scale suggested in this work with 
$\tau_{\rm \star} \sim \tau_{\rm H2}$. 
Indeed, the SF scheme in recent SAMs has been updated to include insights from observations showing that the density of SF correlates strongly with the molecular gas density~\citep[e.g.][]{Blitz:06,Leroy:08}. 
These results have triggered significant activity in SAMs aimed at including an explicit treatment for the transition from atomic to molecular hydrogen, and molecular hydrogen based SF
laws~\citep{Fu:10,Guo:11,Lagos:11a,Somerville:15,Wolz:16}. For example,~\citet{Lagos:11a} and~\citet{Wolz:16}, a modification of the \textsc{SAGE} model presented 
in~\citet{Croton:16}, use a pressure-based transition of neutral to molecular hydrogen~\citep{Blitz:06},
similar to the molecular mass calculation of~\citet{Duffy:12a} that we have adopted in \textsc{Smaug}.
The~\citet{Wolz:16} implementation of \textsc{SAGE} and the SAMs considered in~\citet{Lagos:11a} use a constant consumption time-scale $1/(0.53 \pm 0.25 {\rm Gyr}^{-1}) \approx 2 {\rm Gyr}$, constrained from local Universe observations in~\citet{Leroy:08} and~\citet{Bigiel:11}.

For \textsc{meraxes}, the SFR is $\alpha_{\rm SF} (M_{\rm cold} - M_{\rm crit}) / t_{\rm dyn}$~\citep[][their equation 7]{Mutch:16a}
where discs of $M_{\rm \star}=10^{9}\Msol$ have typical dynamical time-scales $t_{\rm dyn} [\myr] = [149.5, \,14.6,\, 7.9]$ for $z=[0,\,5,\,8]$ and a fiducial SF efficiency
per dynamical time of $\alpha_{\rm SF} = 0.03$ for cold gas mass $M_{\rm cold}$ above a critical threshold $M_{\rm crit}$. This prescription, which follows~\citet{Croton:06, Croton:16}, 
has an effective overall gas consumption time-scale of $\tau_{\rm gas > crit} [\myr] = t_{\rm dyn} / \alpha_{\rm SF} = [4982,\,485,\,265]$ at $z=[0,\,5,\,8]$ which at first glance is approximately 50per cent longer than the $z=5$ time-scale we find in \textsc{Smaug} (Fig.~\ref{fig:f_vs_ssfr}).
However, due to the difference in gas mass definitions, where only gas above the critical density is converted per consumption time, the \textsc{meraxes} gas reservoir 
is even less rapidly used for SF than the simple comparison of time-scales would suggest. By moving to a more explicit molecular
gas mass calculation, rather than cold gas above a critical threshold, SAMs like \textsc{meraxes} could directly use the results from \textsc{Smaug} to more physically model 
SF.\footnote{We note that such a direct comparison between techniques is potentially complicated by baryonic suppression of halo growth when gas physics
is explicitly modelled as shown by~\citet{Qin:17a}. This is particularly an issue for high-redshift dwarf galaxy merger trees generated from $N$-body simulations 
in which halo mass can be significantly overestimated. This direct comparison is further explored in Qin et al. (in preparation).}
The current \textsc{meraxes} formalism, while successfully able to reproduce a range of SFR and stellar masses across redshifts, is a phenomenological relation and it is desirable
to construct a more physical model. 

Based on this discussion, we implement a new SF law informed by~\citet{Blitz:06} to modify \textsc{meraxes} following~\citet{Wolz:16} for three different molecular consumption time-scales.
All other parameters remain as constrained by~\citet{Mutch:16a}. The first two tests are with constant time-scales; a default $2 {\rm Gyr}$ appropriate for low redshifts, 
and $300 \myr$ as suggested by \textsc{Smaug} for high-redshifts. 
We also test a variable consumption time-scale that is a function of redshift and stellar mass according to equation (\ref{eqn:evopowerlaw}) and use the best-fitting models presented in Table~\ref{tab:th2_smass_fits}.
Due to the similarity in the qualitative behaviour of all simulations above \mstarmin in stellar mass, and the relatively fewer systems in the \emph{SSNe} feedback schemes to constrain the fit,
we choose one of the more robustly constrained schemes. In this case, we have tested \emph{ZC\_WSNe\_Kinetic\_EarlyRe} with 
$\tau_{\rm H2} = 190 {\rm Myr} \left[{M_{\star}}/{ 10^{9} \Msol } \right]^{-0.3}\left[ (1+z)/8 \right]^{-1.1}$ that
we note if extrapolated to the present day for low-mass systems at $M_{\star} = 10^{9} \Msol$ gives a similar result of $\sim 2 {\rm Gyr}$ to that found by~\citet{Leroy:08}.

We compare the resulting stellar mass function from the different molecular consumption time-scales in Fig.~\ref{fig:deplt_sam}. It is clear that the shorter time-scale of $300 \myr$ 
recovers the observations of~\citet{Song:16} when compared against the low-redshift calibrated value of $\sim 2 {\rm Gyr}$ and is also in excellent agreement with the 
highly successful, but phenomenological, original \textsc{meraxes} formalism. 
In particular, it is able to match the observed low stellar mass downturn ($M_{\star} < 10^{9} \Msol$) at high-redshift ($z \ge 6$) of~\citet{Gonzalez:11} and~\citet{Duncan:14}.
The shorter time-scale that varies with mass and redshift is required to create an object as massive 
as `GN-z11'~\citep{Oesch:16} at such high-redshift (top left-hand panel of Fig.~\ref{fig:deplt_sam}) as was found in the original \textsc{meraxes} model~\citep{Mutch:16b}.
Therefore, we conclude that adopting such a time-scale represents a marked improvement over longer time-scales set by present-day observations. 
However, we will show through Markov chain Monte Carlo exploration of the parameter space (Mutch et al. in preparation) that the standard \textsc{meraxes} SF law and stellar feedback implementation 
is unable to produce the observed flattened stellar mass function at $z \sim 5$, while still allowing objects such as `GN-z11' to form in time.
We leave the exploration of further observational consequences of the suggested improvement to semi-analytic modelling at high-redshift to future work (Kim et al. in preparation).

\section{Conclusion} \label{sec:conclusion}
During reionization, galaxies are still so young that the recycling of gas or metal enrichment has not had time to significantly impact their growth. 
However high-redshift galaxies, while lacking that complexity, are not in equilibrium and do not grow as expected from the bathtub model. 
We find that early galaxies grow with a molecular gas consumption time-scale of $300 \myr$ at $z\sim 5$ independent of feedback.
The redshift-dependent result, in agreement with simple exponential disc scaling arguments, of $\tau_{\rm H2} = 190 {\rm Myr} \left[{M_{\star}}/{ 10^{9} \Msol } \right]^{-0.3}\left[ (1+z)/8 \right]^{-1.1}$
naturally leads to a time-scale of $2\, {\rm Gyr}$ at $z=0$ in good agreement with local Universe predictions.
It also greatly improves the agreement between SAM predictions and observations of the stellar mass function at higher redshift.

We found that there is no simple relation for the total gas consumption time-scales as gas-rich galaxies can have
sSFRs that differ by two to three orders of magnitude at fixed gas fraction, irrespective of feedback. This suggests that early galaxies were not able to raise
their growth rates as fast as new material arrived. 

In contrast to the standard view of a `booming' early Universe, galaxies of this time were actually in recession, with bottlenecks in transitioning inflowing gas to the 
ISM preventing demand in the galactic economy from responding to the overabundance of supply.
The period of rapid galaxy growth comes later, when galaxies move from demand-side limited recession to a one that is supply-side limited and close to the Universe we live in today.

\section*{Acknowledgements}
ARD would like to thank Caitlin Casey, Katie Mack and Kristian Finlator for stimulating discussions, Claudio Dalla Vecchia for his SNe feedback models, 
as well as Chris Power and Doug Potter for their efforts with initial conditions. JSBW acknowledges the support of an Australian Research Council Laureate Fellowship. 
This work was supported by the Flagship Allocation Scheme of the NCI National Facility at the ANU, data storage at VicNodes, generous allocations of time through the iVEC Partner Share 
and Australian Supercomputer Time Allocation Committee. ARD gratefully acknowledges the use of computer facilities purchased through a UWA Research Development Award. 
We would like to thank the \textsc{Python} developers of \textsc{matplotlib}~\citep{Hunter:07}, \textsc{CosmoloPy} (http://roban.github.com/CosmoloPy) and 
\textsc{pynbody}~\citep{pynbody} for easing the visualization and analysis efforts in this work.

\appendix
\section{Resolution Testing}\label{sec:appendix}
We have extensively explored resolution tests for the \textsc{Smaug} simulation series, focusing on comparing the increasingly well-resolved models
with $N^3$ gas and $N^3$ DM particles in the same initial conditions of cubic volumes of comoving length $L = 10\hMpc$. 
In our tests, we explored $N=[128,256,512]$ resulting in Plummer-equivalent comoving softening lengths of $[200, 400, 800 \hpc]$ 
and DM (gas) particle masses of $[4.7\, (0.9),38\, (7.6),300\, (61)] \times 10^{5} \hMsol$, respectively.

\begin{figure}
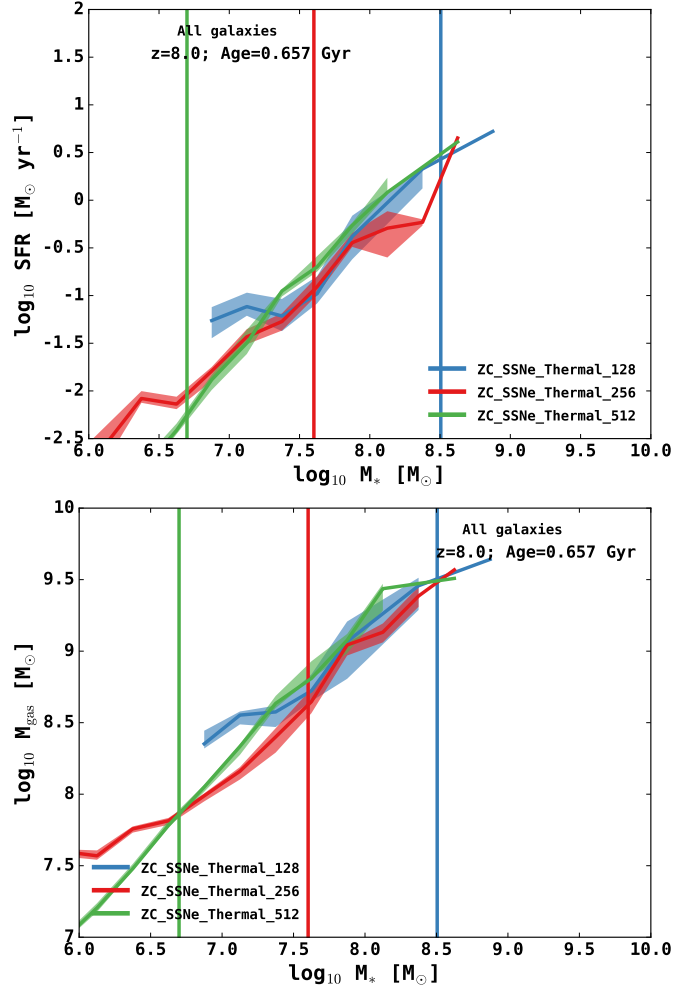

\centering
\begin{subfigure}{\red\textwidth}
\includegraphics[width=\textwidth,keepaspectratio=true]{{{sfr_vs_smass_z\res_Res}}}
\end{subfigure}
\begin{subfigure}{\red\textwidth}
\includegraphics[width=\textwidth,keepaspectratio=true]{{{gmass_vs_smass_z\res_Res}}}
\end{subfigure}
\caption{Here, we consider the resolution tests of the median SFR (gas mass) as a function of stellar mass in the top (bottom) panel for three increasingly
well-resolved simulations for the maximal feedback physics scheme \emph{ZC\_SSNe\_Thermal}. The vertical lines are the estimated
resolution limits for stellar mass and indicate that the quantities in the vertical axis are well behaved above these conservative limits.}\label{fig:restest_smass}
\end{figure}

We explored mass functions for all our basic properties and found that baryonic processes, $V$, were resolved at high-redshift above 
\begin{equation}\label{eqn:reslaw}
V_{\rm min} = V_{\rm lim} \left(\frac{L}{10\hMpc} \frac{512}{N}\right)^{3}\,,
\end{equation}
where $V_{\rm lim}$ depends on the variable of interest. Stellar, gas and molecular masses have $V_{\rm lim} = [5, 50, 50] \times 10^{6} \hMsol$
respectively while $\sSFR_{\rm lim} = 2\times 10^{-2} \gyr^{-1}$. 
We also found that a conservative cut on halo mass of $M_{\rm halo, \rm lim}=10^{8.5} \hMsol$ resulted in most derived quantities for those larger
haloes being consistent. 

We show an example resolution test in Fig.~\ref{fig:restest_smass} with the SFR (gas mass) shown against stellar mass in the top (bottom) panel.
The vertical lines are the cuts from equation (\ref{eqn:reslaw}) and show that the systems are well behaved above these limits.

\begin{figure}
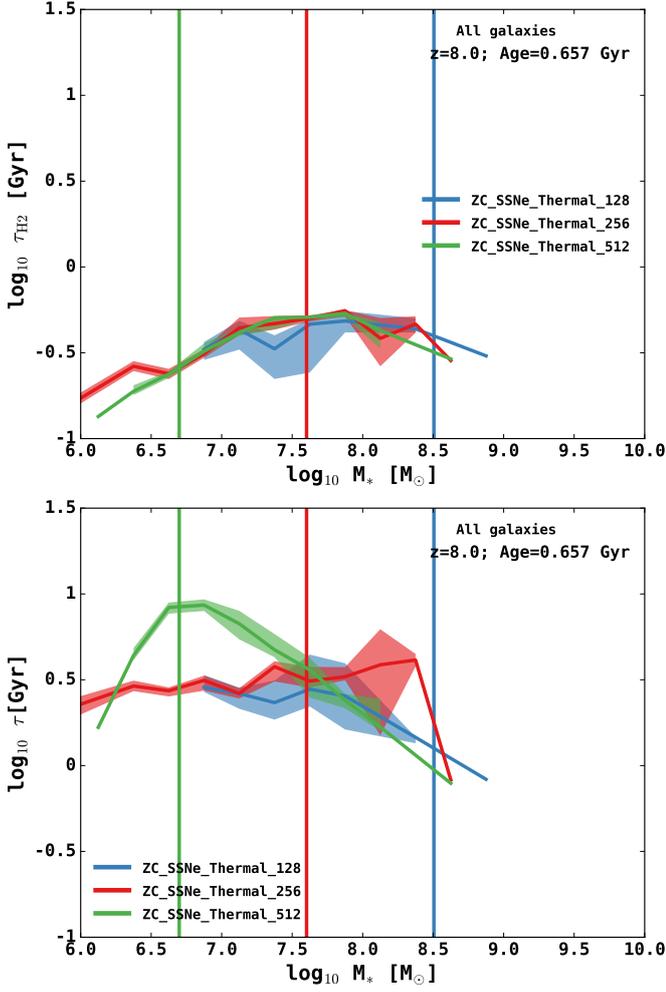

\centering
\begin{subfigure}{\red\textwidth}
\includegraphics[width=\textwidth,keepaspectratio=true]{{{th2_vs_smass_z\res_Res}}}
\end{subfigure}
\begin{subfigure}{\red\textwidth}
\includegraphics[width=\textwidth,keepaspectratio=true]{{{tau_vs_smass_z\res_Res}}}
\end{subfigure}
\caption{Here, we consider the resolution tests of the molecular (total gas) consumption time-scale as a function of stellar mass in the top (bottom) panel for three increasingly
well-resolved simulations for the maximal feedback physics scheme \emph{ZC\_SSNe\_Thermal}. The vertical lines are the estimated
resolution limits for stellar mass and indicate that the quantities in the vertical axis are well behaved above these conservative limits. It is apparent that 
the turnover in the molecular consumption time-scale is not caused by limited numerical sampling, while the total gas consumption time-scale 
is more sensitive to resolution.}\label{fig:restest_tau}
\end{figure}

It is more difficult to determine the exact numerical convergence, or mass cut-off, for the other derived quantities explored in this work. 
In Fig.~\ref{fig:restest_tau}, we see that the characteristic turnover of the consumption time-scale at low stellar mass (left column of Fig.~\ref{fig:th2_deplt}) 
appears to be independent of resolution. 
The overall gas consumption time-scale (bottom panel) suffers significantly more obviously from numerical issues; however, the minimum stellar 
halo mass we consider for our fits of \mstarmin is well above even the most conservative of estimates.

\begin{figure}
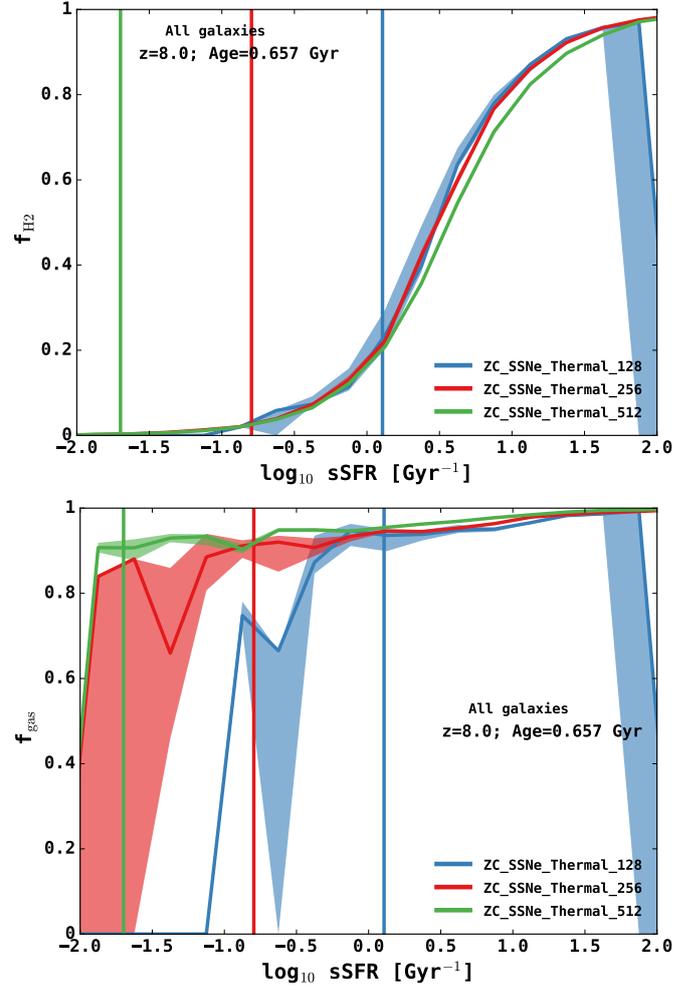

\centering
\begin{subfigure}{\red\textwidth}
\includegraphics[width=\textwidth,keepaspectratio=true]{{{fh2_vs_ssfr_z\res_Res}}}
\end{subfigure}
\begin{subfigure}{\red\textwidth}
\includegraphics[width=\textwidth,keepaspectratio=true]{{{fgas_vs_ssfr_z\res_Res}}}
\end{subfigure}
\caption{Here, we consider the resolution tests of the molecular (total gas) fraction time-scale as a function of the sSFR
in the top (bottom) panel for three increasingly well-resolved simulations for the maximal feedback physics scheme \emph{ZC\_SSNe\_Thermal}. 
The vertical lines are the estimated resolution limits. The total gas fraction has a well-modelled numerical resolution effect, whereas the molecular
gas fraction is less obviously affected. This pattern is similar to the related consumption time-scale plots shown in Fig.~\ref{fig:restest_tau}.}\label{fig:restest_ssfr}
\end{figure}

As we see in Fig.~\ref{fig:restest_ssfr} the total gas fraction (bottom panel) has a more obvious numerical resolution limit than that of the molecular gas (top panel).
The increasing resolutions allow the gas fraction to be sampled at ever lower sSFR, with scaling given by equation (\ref{eqn:reslaw}). The case for the molecular
gas fraction (as with Fig.~\ref{fig:restest_tau}) appears to be less obviously affected by resolution. This agreement should not be overly surprising as the consumption
time-scale is the common variable between these two tests.

\label{lastpage}
\end{document}